\documentclass[a4paper,11pt]{article}
\pdfoutput=1 

\usepackage{jheppub}

\usepackage[T1]{fontenc} 
\usepackage{graphicx}

\usepackage{amsmath}
\usepackage{amssymb}
\usepackage{epsfig}
\usepackage{graphicx}
\usepackage{slashed}
\usepackage{bm}
\usepackage{soul}
\usepackage{url}
\usepackage[normalem]{ulem} 
\usepackage{color}

\begin{document}

\title{Bubble Expansion and the Viability of Singlet-Driven Electroweak Baryogenesis}

\author[]{Jonathan Kozaczuk}

\affiliation[]{TRIUMF,\\4004 Wesbrook Mall,\\ Vancouver, BC V6T 2A3, Canada }

\emailAdd{jkozaczuk@triumf.ca}

\abstract{The standard picture of electroweak baryogenesis requires slowly expanding bubbles. This can be difficult to achieve if the vacuum expectation value (VEV) of a gauge singlet scalar field changes appreciably during the electroweak phase transition.  It is important to determine the bubble wall velocity in this case, since the predicted baryon asymmetry can depend sensitively on its value.  Here, this calculation is discussed and illustrated in the real singlet extension of the Standard Model. The friction on the bubble wall is computed using a kinetic theory approach and including hydrodynamic effects.  Wall velocities are found to be rather large ($v_w \gtrsim  0.2$) but compatible with electroweak baryogenesis in some portions of the parameter space. If the phase transition is strong enough, however, a subsonic solution may not exist, precluding non-local electroweak baryogenesis altogether. The results presented here can be used in calculating the baryon asymmetry in various singlet-driven scenarios, as well as other features related to cosmological phase transitions in the early Universe, such as the resulting spectrum of gravitational radiation.}

\maketitle
\flushbottom

\setcounter{page}{2}
\section{Introduction}

Cosmological phase transitions in the early Universe are interesting for a variety of reasons. They can produce observable gravitational radiation \cite{Turner:1990rc, Hogan:1986qda, Witten:1984rs, Kamionkowski:1993fg}, seed primordial magnetic fields~\cite{Turner:1987bw, Baym:1995fk}, affect the abundance of thermal relics \cite{Cohen:2008nb, Wainwright:2009mq}, and otherwise play an important role in the cosmological history of the Universe~\cite{Boyanovsky:2006bf}. Perhaps most notably, such phase transitions can give rise to a viable mechanism of baryogenesis, provided the transition is first-order.

A first-order phase transition can occur in the early Universe when two vacua of the theory coexist for some range of temperatures. If this is the case, an energy barrier exists between the two and the system can transition to the state with lower free energy via quantum tunneling or thermal fluctuations \cite{Coleman:1977py, Callan:1977pt, Linde1, Linde2}. Physically, this corresponds to the formation of spherical bubbles in the ambient metastable vacuum. These bubbles grow and can reach a steady state expansion velocity. Inside the bubble, some subset of the scalar fields have non-zero vacuum expectation values (VEVs), while outside they do not. If the non-vanishing condensate breaks the $SU(2)_L$ gauge symmetry of the Standard Model (SM), as at the electroweak phase transition (EWPT), non-perturbative sphaleron transitions, which violate $B+L$, will be quenched inside the bubble and active outside. These processes can act on chiral charge currents diffusing in front of the wall to source a net baryon asymmetry. If the sphaleron rate is significantly suppressed in the broken electroweak phase, the asymmetry can be frozen in once captured by the expanding bubble. Roughly, this requires~\cite{Quiros_review, Patel:2011th,Fuyuto:2014yia}
\begin{equation}
\frac{\langle h \rangle}{T}\gtrsim 1,
\end{equation}
where $T$ is a temperature associated with the phase transition and $h$ is some combination of fields charged under $SU(2)_L$ usually identified with the Standard Model-like Higgs. This picture is known as "non-local" or "transport-driven" electroweak baryogenesis (EWB) and is an elegant explanation for the origin of the observed baryon asymmetry of the Universe \cite{Cohen:1990it, Cohen:1990py, Nelson:1991ab, Joyce:1994zn, Joyce:1994zt, Morrissey:2012db}.

Clearly this mechanism relies on several requirements beyond sphaleron suppression in the broken phase. One must ensure a significant amount of $CP$-violation to source the chiral charge currents in front of the wall. This is well-known, has been studied extensively, and has motivated several experiments in search of $CP$-violating signatures, such as permanent electric dipole moments \cite{Engel:2013lsa}. There is another, somewhat less-appreciated, requirement for successful baryogenesis through this mechanism: bubbles must expand slowly enough for sphalerons to convert a significant fraction of the $CP$-asymmetry to a net baryon density~\cite{Nelson:1991ab, Joyce:1994fu, Joyce:1994zn, Joyce:1994zt}. In other words, the diffusion of the chiral plasma excitations must be efficient in front of the bubble wall\footnote{Bubbles must also not expand too slowly; otherwise a quasi-equilibrium situation is reached and the net baryon density is equilibrated away. This is typically not a problem in baryogenesis scenarios, since it requires bubbles moving very slowly, with $v_w\lesssim 0.01$ \cite{Nelson:1991ab}}.

Precisely how slowly the wall must move in this scenario depends on several factors, including the amount of $CP$-violation and the details of diffusion in front of the bubble.  However, it is generally the case that bubble expansion must at least be slower than the speed of sound in the plasma\footnote{More precisely, it is the wall velocity relative to the fluid in front of the bubble that should be subsonic~\cite{No:2011fi}. However, in what follows the fluid velocity in the symmetric phase will always be perturbatively small, and so we will simply require the wall velocity to be subsonic in the rest frame of the fluid far from the bubble. For further discussion on this point, see Ref.~\cite{No:2011fi}. }, $c_s\sim 1/\sqrt{3} \approx 0.58$. Otherwise, diffusion in front of the wall will be very inefficient. Even if the wall moves subsonically, the predicted value of the steady-state wall velocity, $v_w$, is an important input into any non-local electroweak baryogenesis calculation. Previous studies have found that the predicted baryon asymmetry typically peaks around $v_w\sim 0.01$, falling off as $\sim1/v_w$ or faster for larger values~\cite{Heckler:1994uu, Huber:2001xf, Carena:2000id, Kozaczuk:2011vr} (for velocities much smaller than $0.01$, the sphalerons can begin to equilibrate the asymmetry). In some cases the dependence of the predicted asymmetry on the wall velocity can be less severe~\cite{Konstandin:2005cd, Huber:2006wf, Huber:2006ma}, as the scaling hinges on the form of the dominant $CP$-violating source (see e.g. Refs.~\cite{Kainulainen:2002th, Prokopec:2003pj,  Konstandin:2004gy}).  Nevertheless, a determination of the wall velocity is often important even for rough estimates of the baryon asymmetry when the wall is expected to move quickly.

The electroweak phase transition in the Standard Model is not first-order \cite{Kajantie:1995kf, Kajantie:1996mn}. Thus, electroweak baryogenesis necessarily requires some physics beyond the SM. There have been many such scenarios proposed in the literature. One of the most popular and straightforward involves augmenting the Standard Model Higgs sector by a gauge singlet scalar field \cite{Pietroni:1992in, Davies:1996qn, Huber:2000mg, Menon:2004wv, Ham:2004cf,  Profumo:2007wc, Bodeker:2009qy, Espinosa:2011ax, Kozaczuk:2013fga, Huang:2014ifa, Kozaczuk:2014kva, Profumo:2014opa, Curtin:2014jma, Jiang:2015cwa}, providing a tree-level cubic term in the effective potential. Such a cubic term can easily give rise to the barrier required for a first-order transition, and the singlet nature of the new state(s) in many cases can ensure compatibility with current phenomenological constraints~\cite{Huang:2014ifa, Kozaczuk:2014kva, Profumo:2014opa, Curtin:2014jma, Jiang:2015cwa}, including the observation of the 125 GeV Standard Model-like Higgs at the LHC~\cite{Aad:2012tfa, Chatrchyan:2012ufa}. This class of models, though popular and simple, is often expected to produce fast-moving bubble walls when the singlet field VEV changes appreciably during the electroweak phase transition \cite{Bodeker:2009qy}. This is simply because the additional field direction contributes to the pressure difference between the phases, which drives the expansion of the bubble, but does not experience a substantial drag force from the plasma\footnote{If the singlet field is instead approximately stabilized during the electroweak phase transition, bubbles can expand significantly more slowly~\cite{Konstandin:2014zta}).}. Although this fact was recognized several years ago~\cite{Bodeker:2009qy}, studies of EWB typically focus on the strength of the phase transition and assume $CP$-violation can be added in separately without considering the effects of the bubble wall dynamics on generating the baryon asymmetry.  In fact, while many studies have since considered baryogenesis in these scenarios with a changing singlet VEV, the wall velocity has never been directly calculated\footnote{In Ref.~\cite{Kozaczuk:2014kva}, we performed a very rough estimate of the wall velocity in the NMSSM. Several important terms in the Boltzmann equations were dropped, likely resulting in a significant under-estimate of the wall velocity. A full microphysical calculation of $v_w$ in the NMSSM does not currently exist in the literature. Such a study can be undertaken with the methods discussed here and is currently in progress.}, nor has it been shown that the resulting bubble walls can propagate subsonically as required for successful EWB. Our aim here is to fill this gap.  

The bubble wall velocity is an important quantity to compute apart from baryogenesis considerations. For example, models with additional gauge singlets that predict a strong first-order transition can source gravitational waves through bubble collisions and turbulence (see e.g. Refs.~\cite{Turner:1990rc, Kosowsky:1991ua, Kosowsky:1992rz, Kosowsky:1992vn, Kamionkowski:1993fg, Kosowsky:2001xp, Caprini:2006jb, Gogoberidze:2007an, Huber:2007vva, Caprini:2007xq, Huber:2008hg, Caprini:2009fx, Caprini:2009yp, Leitao:2012tx, Hindmarsh:2013xza, Hindmarsh:2015qta}). These scenarios may be effectively probed by upcoming gravitational wave experiments, such as eLISA~\cite{ Binetruy:2012ze}, or Big Bang Observer~\cite{Corbin:2005ny}. However, in order to connect these observations to an underlying theory, one must be able to reliably calculate the velocity of the expanding bubble, as well as the other properties of the phase transition.

A detailed calculation of the wall velocity is rather involved. Building on previous work~\cite{Enqvist:1991xw, Dine:1992wr, Liu:1992tn, Ignatius:1993qn, Carrington:1993ng, Heckler:1994uu}, Moore and Prokopec were the first to calculate the velocity for the Standard Model case microphysically in Refs.~\cite{Moore:1995ua,  Moore:1995si} (Ref.~\cite{Konstandin:2014zta} also recently revisited this calculation). Five years later, John and Schmidt performed an analogous study in the minimal supersymmetric Standard Model (MSSM) with a light scalar top quark (stop) \cite{John:2000zq}. Around the same time, the effects of infrared gauge bosons on the wall velocity were calculated in Ref.~\cite{Moore:2000wx}.  Most recently, Ref.~\cite{Konstandin:2014zta} extended the results of Moore and Prokopec to other SM-like scenarios, including that in which the VEV of a singlet scalar is approximately stabilized during the EWPT. To date, these remain the only full microphysical calculations of the wall velocity existing in the literature\footnote{By `microphysical', we mean calculations explicitly computing the friction exerted by the plasma on the wall, as opposed to those using a phenomenological viscosity parameter. The former involves determining the various interaction rates in the electroweak plasma and solving for the deviations from thermal equilibrium around the bubble wall, as we describe in detail below.}.  Recent years have seen progress in matching models onto these existing results~\cite{ Megevand:2009gh, Huber:2011aa, Huber:2013kj, Konstandin:2014zta} and in the hydrodynamic considerations associated with bubble wall expansion (see e.g.~\cite{Espinosa:2010hh, Leitao:2010yw, Megevand:2013hwa, Megevand:2013yua, Leitao:2014pda,  Megevand:2014yua, Megevand:2014dua}). 

Scenarios in which an the VEV of an additional singlet scalar field changes appreciably during the transition merit separate consideration\footnote{In the remainder of this study we will take `singlet-driven scenarios' to refer to those in which the singlet VEV changes non-negligibly during the transition. This can occur even in models with a discrete $\mathbb{Z}_2$ symmetry at $T=0$. Examples in which the singlet VEV is approximately stabilized during the EWPT can be treated by the techniques developed for the Standard Model (or MSSM, if the singlet contributions to the finite-$T$ cubic term are large) since only the Higgs field is involved in the transition (see e.g. Refs.~\cite{Huber:2013kj, Konstandin:2014zta} for an application of this approach).}. This is because one must account for the friction on the singlet field. Neglecting these contributions can lead to a drastic over-estimate of the wall velocity. A proper treatment requires computing several new classes of interaction rates in the plasma, which can be rather involved. Also, the additional field direction complicates the equations of motion for the condensates. Nevertheless, the calculation can be done and is especially important given the current status of electroweak baryogenesis in light of collider searches.  Until recently, the MSSM light stop scenario ~\cite{Carena:1996wj, Delepine:1996vn} was considered by many to be the most plausible setting for electroweak baryogenesis beyond the SM. Now light stops are in severe tension with both direct LHC searches~\cite{Delgado:2012eu, Krizka:2012ah} and measurements of the Standard Model-like Higgs couplings~\cite{Cohen:2012zza, Menon:2009mz, Curtin:2012aa, Carena:2012np}. Similar conclusions hold true for many different models relying on large thermally-induced cubic terms to strengthen the phase transition~\cite{Cohen:2012zza, Chung:2012vg, Katz:2014bha}. This situation has led to a renewed interest in singlet-driven scenarios, since they can be much more elusive at colliders~\cite{Curtin:2014jma}.  An analysis of the wall dynamics would mark an important step forward in understanding electroweak baryogenesis in these models. 

The goal of this study will be to demonstrate how the electroweak bubble wall velocity can be calculated and to extract some general features of the wall dynamics in singlet-driven scenarios . We will generally follow the strategy and techniques developed in Refs.~\cite{Moore:1995ua,  Moore:1995si, Moore:2000wx, John:2000zq, Konstandin:2014zta} but modified to account for the singlet field direction. The methods described will be applicable to many different models, although for the sake of simplicity we will frame our discussion in the real singlet extension of the SM, sometimes known as the `xSM' \cite{Profumo:2007wc}.This scenario should encapsulate the most relevant features of models with singlet-driven first-order phase transitions.  Notably, the xSM does not feature any new sources of $CP$-violation, which could in principle significantly alter the wall dynamics when included. We comment further on this below and the reader should keep this in mind as we proceed. 

We will focus on two different schemes for calculating the wall velocity. The first is explicitly gauge-independent and neglects the contributions to the effective potential and friction from the $SU(2)_L$ gauge bosons, while the second includes them. For slower bubble walls (such that the friction on the singlet field is large), both calculations yield similar results, while for faster walls the gauge boson contributions become increasingly important in slowing down the expansion.

The remainder of this study is structured as follows. In Section~\ref{sec:xSM}, we introduce the singlet-driven scenario and the finite temperature effective potential which will be used throughout our study. We then move on to computing the wall velocity. Following Ref.~\cite{Moore:1995si}, the calculation can be broken down into several parts. First, the phase transition properties must be computed and the temperature near the bubble inferred from hydrodynamic considerations, as discussed in Sec.~\ref{sec:hydro}. There we also discuss the equations of motion (EOMs) for the bubble wall and consider the simple case of wall velocities in the ultra-relativistic limit. Possible values for the steady state wall velocity are those such that the equations of motion for the wall are satisfied. The EOMs depend on the deviations from thermal equilibrium of all plasma excitations in front of the wall. These are discussed in Sec.~\ref{sec:Boltz}. We next move on to solving the system of equations for the deviations from equilibrium and the equations of motion in Sec.~\ref{sec:solve}. The calculation is then applied to the parameter space of the xSM consistent with all phenomenological constraints in Sec.~\ref{sec:resxSM}. We find that sufficiently strong phase transitions may possess no subsonic solutions, and that $v_w\gtrsim 0.2$ for points with $\langle h \rangle/T_n\geq 1$ in the parameter regions considered. We infer that bubbles may expand slowly enough for singlet-driven electroweak baryogenesis, but only in certain portions of the parameter space. Results for the bubble wall profiles are also presented. Finally, our main findings and conclusions are summarized in Sec.~\ref{sec:summ}. We also provide a brief appendix which compares the interaction rates we have calculated with those appearing elsewhere in the literature.

Before proceeding it is important to note that there are other electroweak baryogenesis scenarios that do not rely on diffusion in front of the wall, and hence that do not require slow bubble walls. Local EWB~\cite{ Turok:1990in, McLerran:1990zh, Dine:1990fj, Lue:1996pr}, in which the baryon number and $CP$-violation occur in the same region at the bubble wall boundary, is one such example\footnote{Cold Electroweak Baryogenesis~\cite{Tranberg:2003gi, Tranberg:2009de, Konstandin:2011ds} also falls into this category.}. However, this typically leads to a highly suppressed total baryon asymmetry relative to the non-local case, since the sphaleron transitions turn off  near the outer edge of the bubble wall~\cite{Nelson:1991ab}. More recently, an interesting scenario was presented in Ref.~\cite{Caprini:2011uz}, in which bubbles can expand quickly enough to significantly reheat the plasma inside the bubble. Secondary bubbles can then nucleate near which transport-driven baryogenesis can occur. This is an intriguing possibility,  however it requires a substantial amount of reheating which only occurs for very strong phase transitions. The reader should bear these alternative scenarios in mind as we proceed. 

\section{A Singlet-Extended Higgs Sector}\label{sec:xSM}

To study the dynamics of singlet-driven electroweak phase transitions, we will work in the real singlet extension of the Standard Model. This simple scenario has been studied in depth in the literature, from the standpoint of electroweak baryogenesis, dark matter, LHC signatures, and more (see e.g. Refs.~\cite{Ham:2004cf, O'Connell:2006wi, Profumo:2007wc, Barger:2007im, He:2009yd, Gonderinger:2009jp, Espinosa:2011ax, Cline:2012hg, Cline:2013gha, Profumo:2014opa} and references therein). The arguments and methods we discuss here can be straightforwardly applied to other singlet extensions of the SM, such as the next-to-minimal supersymmetric Standard Model (nMSSM~\cite{Menon:2004wv, Huber:2006wf, Huber:2006ma} or NMSSM~\cite{Pietroni:1992in, Davies:1996qn, Huber:2000mg, Huber:2006ma, Huang:2014ifa, Kozaczuk:2014kva}) and other scenarios with real or complex gauge singlets \cite{Jiang:2015cwa}. 

The tree-level potential is taken to be\footnote{One is free to shift the singlet field value such that the $T=0$ tadpole is removed~\cite{Espinosa:2011ax}.} 
\begin{equation}
\begin{aligned}
V_0(H,S)=&-\mu^2(H^{\dagger}H)+\lambda (H^{\dagger}H)^2+\frac{1}{2}a_1 (H^{\dagger}H) S+\frac{1}{2}a_2 (H^{\dagger}H)S^2\\&+\frac{1}{2}b_2 S^2+\frac{1}{3}b_3 S^3+\frac{1}{4} b_4 S^4
\end{aligned}
\end{equation}
where $S$ is a real scalar singlet under the Standard Model gauge groups and $H$ is a complex $SU(2)_L$ doublet. Both the singlet and $CP$-even neutral component of $H$ are assumed to obtain vacuum expectation values during electroweak symmetry breaking. Throughout our discussion, we will also assume that both VEVs vanish in the high-temperature phase for simplicity (this is discussed further below). Non-vanishing VEVs correspond to minima of the effective potential for non-zero background field values $\phi_h$, $\phi_s$. These classical fields are those relevant for computing the properties of the phase transition.  At a given temperature, we can expand $H$ and $S$ about the background fields,
\begin{equation}
H^{\rm T}=\left(
\phi^+,
\frac{\phi_h(T)+h+i \phi^0}{\sqrt{2}}
\right),
\hspace{0.3cm}
S=\phi_s(T)+s.
\end{equation}
At zero temperature, $\phi_h(T=0)\equiv v=246$ GeV. The zero temperature singlet VEV, $\phi_s(T=0)\equiv v_s$ can vary. 
 
Throughout our study we will identify $h$ with the Standard Model-like Higgs discovered at the LHC~\cite{Aad:2012tfa, Chatrchyan:2012ufa} and take $s$ to be a pure singlet with no mixing at tree-level. The phenomenology of this setup and our choices for the various parameters are detailed in Section~\ref{sec:pheno} below. 

\subsection{The Effective Potential}

For a homogeneous background field configuration $\phi(x)\equiv \phi$, the ground state of the theory corresponds to a minimum of the effective potential $V_{\rm eff}(\phi)$.  At one loop, $V_{\rm eff}$ is given by the tree-level potential (expanded around the background fields), modified by additional Coleman-Weinberg terms. 

At finite temperature and density the physical ground state of the theory is altered by the interactions of the scalar field $\phi$ with the ambient plasma. The vacua of the theory can then be determined from the finite-temperature effective potential, $V_{\rm eff}(\phi, T)$. In the simple case involving one background field, it is given by~\cite{Quiros_review}
 \begin{align}\label{eq:VT}
 V^T(\phi,T)=\frac{T^4}{2\pi^2}\left[\sum_i \pm N_i J_{\pm}\left(\frac{m^2_i(\phi)}{T^2}\right)\right],
 \end{align}
 where the plus and minus signs correspond to the bosonic and fermionic contributions, respectively,  and the $N_i$ are the associated number of degrees of freedom for the species $i$.  This expression generalizes straightforwardly to the case of more than one background field.  The functions $J_{\pm}$ are given by
 \begin{equation}
  J_{\pm}(x)=\int_0^\infty dy\, y^2\log\left[1\mp \exp(-\sqrt{x^2+y^2})\right].
 \end{equation}
In the high-temperature limit they admit a useful expansion, given by
\begin{align}
\label{eq:J}
 T^4 J_+\left(\frac{m^2}{T^2}\right)=&-\frac{\pi^4 T^4}{45}+\frac{\pi^2m^2 T^2}{12}-\frac{T\pi (m^2)^{3/2}}{6}-\frac{(m^4)}{32}\log\frac{m^2}{a_b T^2},\\
 \nonumber T^4 J_-\left(\frac{m^2}{T^2}\right)=&\frac{7\pi^4 T^4}{360}-\frac{\pi^2m^2 T^2}{24}-\frac{(m^4)}{32}\log\frac{m^2}{a_f T^2}, 
\end{align}
with $a_b=16\pi^2e^{3/2-2\gamma_E},\,a_f=\pi^2e^{3/2-2\gamma_E}$, and $\gamma_E$ the Euler-Mascheroni constant.  Note that the thermal contributions above correspond to momentum integrals of equilibrium distribution functions for all species in the plasma coupled to $\phi$ ~\cite{Quiros_review, Bodeker:2009qy}. 

\subsection{Gauge-Invariance}\label{sec:gauge_inv}

The finite temperature effective potential is only gauge-invariant at its extrema \cite{  Nielsen:1975fs, Fukuda:1975di}. Thus, tunneling calculations depending on the potential away from the local minima are in general gauge-dependent. This will result in a gauge-dependent determination of the nucleation temperature for the phase transition, $T_n$, and ultimately the wall velocity. To avoid this as much as possible, our primary analysis will only consider terms in the effective potential which are explicitly gauge-invariant. Thus, we will not include the $T=0$ Coleman-Weinberg corrections, or the finite temperature cubic and tadpole terms in the high-temperature effective potential (gauge-dependence in the tadpole may enter at higher perturbative order \cite{Profumo:2014opa}). This is precisely the strategy followed by Ref.~\cite{Profumo:2014opa} in analyzing the phase transition properties of the xSM. The finite-temperature effective potential in this case becomes
\begin{equation}\label{eq:pot}
\begin{aligned}
V_{\rm eff}(\phi_h,\phi_s,T)\simeq &-\frac{1}{2}\mu^2 \phi_h^2+\frac{1}{4}\lambda \phi_h^4 +\frac{1}{4} a_1 \phi_s \phi_h^2+\frac{1}{4}a_2 \phi_h^2\phi_s^2+\frac{1}{2} b_2 \phi_s^2 +\frac{1}{3} b_3 \phi_s^3+\frac{1}{4}b_4 \phi_s^4\\
& + \frac{\phi_h^2 T^2}{96} \left(9 g_2^2+3g_1^2+12y_t^2+24 \lambda +2 a_2\right)+\frac{\phi_s^2 T^2}{24} \left(2 a_2+3 b_4\right).
\end{aligned}
\end{equation}

Although morally satisfying, the gauge-invariant approach has the disadvantage of sometimes neglecting numerically important contributions to the effective potential. While it should capture the physics we are interested in, namely the singlet contributions to the potential and the friction on the expanding bubble, it is important to consider the effects of the neglected terms, especially the gauge boson cubic term\footnote{The inclusion of the tadpole term acts primarily to shift the high-temperature minimum away from $\phi_s=0$. We have computed the wall velocities for several scenarios with this term included and obtain values similar to those found neglecting the tadpole. For simplicity we will not consider this term further in this study, although our methods can be straightforwardly modified to include it.}. To this end, we will also show results for the wall velocities including the gauge boson cubic term and friction in Landau gauge. In this case the effective potential in Eq.~\ref{eq:pot} is modified by the additional term
\begin{equation} \label{eq:cubic}
 \Delta V_{\rm eff}^{\rm cubic} (\phi_h,T) \simeq - \frac{\phi_h^3 T}{12\pi}\left[ \frac{3}{4} g_2^3+\frac{3}{8}\left(g_2^2+g_1^2\right)^{3/2}\right].
 \end{equation}
We will find that these contributions (and the friction from gauge bosons) are numerically significant in some cases, especially for faster moving bubble walls.

\section{Preliminaries: Phase Transitions, Hydrodynamics, and the Wall Equations of Motion} \label{sec:hydro}

For a given point in the model parameter space with a first-order electroweak phase transition, we are interested in determining the steady-state velocity of the bubble wall separating the electroweak-symmetric and broken phases. To do so we must first determine the properties of the phase transition (most importantly its characteristic temperature), as well as the equations of motion for the scalar field condensates. Our discussion will roughly follow that of Ref.~\cite{Moore:1995si}, which we draw from frequently throughout the remainder of this study. 

\subsection{Bubble Nucleation}
Using the effective potential in Eq.~\ref{eq:pot}, first-order transitions can occur when two local minima coexist for some range of temperatures.  The background fields can then ``tunnel" from the origin to the new vacuum, in which $\phi_h$, $\phi_s\neq 0$. This can begin to occur  below the critical temperature, $T_c$, at which the two relevant vacua are degenerate. Bubbles begin to nucleate efficiently at the nuceation temperature, $T_n$,  determined by requiring the expectation value for one bubble to nucleate per Hubble volume to be $\sim\mathcal{O}(1)$.  At finite temperature, the nucleation probability is determined by the $O(3)$-symmetric instanton interpolating between the metastable and and true vacua with the lowest ratio of three-dimensional Euclidean action\footnote{For all the cases we consider, the nucleation temperatures are much larger than the inverse radii of the instantons, and so the $O(3)$ bounce is indeed the relevant quantity to consider. } to temperature, $S_3/T$ \cite{Linde1, Linde2, Quiros_review}.  The tunneling probability per unit volume is then given by
\begin{equation}
\frac{\Gamma}{V}=A(T) e^{-S_3/T}.  
  \end{equation}
where the pre-factor $A(T)$ is only weakly temperature-dependent.  Using dimensional analysis to estimate $A(T)$ and assuming typical electroweak temperatures, one finds that the nucleation temperature is approximately determined by $S_3(T_n)/T_n\approx 140$ \cite{Quiros_review}. We adopt this definition in the rest of our study.

Inside the nucleated bubble, both the VEV of the Higgs and singlet fields will be non-zero by assumption. From the standpoint of electroweak baryogenesis, only the value of $\phi_h$ is important for sphaleron suppression. We will  therefore define a strongly first-order phase transition, occurring at the nucleation temperature $T_n$, by~\cite{Fuyuto:2014yia}
 \begin{equation}
 \frac{\phi_h (T_n)}{T_n}\gtrsim 1.
 \end{equation}
This condition is free from explicit gauge-dependence in our primary setup, since we have neglected all 
 gauge-dependent terms in the effective potential. It should be noted that the baryon number preservation condition above still contains several implicit assumptions, discussed in detail in Ref.~\cite{Patel:2011th}. 
 
\subsection{Temperature Variations}

The nucleation temperature defined above is that of the ambient plasma when bubbles begin to form efficiently. Once formed, however, the temperature is no longer homogeneous. The phase transition releases latent heat into the plasma and the expansion of a subsonic bubble heats up the medium in front of it. The temperature in the broken phase will thus differ from that immediately outside the bubble, which in turn is not the same as the typical nucleation temperature of the bubble.  To relate these various quantities requires a treatment of the plasma hydrodynamics.  These changes in temperature can have large effects on the expansion of the bubble \cite{Moore:1995si}, and so we must take them into account.

Far away from the bubble, the relevant temperature is that at which bubble nucleation occurs, $T_n$.  We wish to obtain the temperature in the vicinity of bubble wall. To do so, let us consider the wall-plasma system, with the plasma modeled as a perfect relativistic fluid. Hydrodynamic equations can be obtained by requiring conservation of the wall-fluid stress-energy tensor~\cite{Enqvist:1991xw},
\begin{equation}\label{eq:fluid}
\partial_{\mu} T^{\mu \nu} = \partial_{\mu} T^{\mu \nu}_{\rm condensate}+\partial_{\mu}T^{\mu \nu}_{\rm plasma} = 0.
\end{equation}
We define the `fluid --' or `plasma frame' such that the fluid is at rest far from the bubble and in its center. This is the frame which we use to define the wall velocity $v_w$ and the wall profile parameters. Solutions to the fluid equations in the plasma frame can typically be classified as either `detonations', in which the bubble velocity exceeds the sound speed $c_s$ in the plasma, or `deflagrations', in which $v_w<c_s$~\footnote{There are also `hybrid' cases; see Ref.~\cite{Espinosa:2010hh}.}.  Successful subsonic electroweak baryogenesis typically requires a deflagration solution, since otherwise diffusion in front of the bubble is inefficient. We will restrict ourselves to this case. 

Consider an expanding bubble with free energy $V_{\rm eff}(\phi_{-},T_-)$ inside and $V_{\rm eff}(\phi_+,T_+)$ immediately outside (`$\pm$' subscripts will correspond to quantities outside/inside the bubble).  The equations of state (EoS) for the two phases can be written as
\begin{equation}
p_{\pm}=\frac{1}{3}a_{\pm} (T) T_{\pm}^4-\epsilon_{\pm}(T), \hspace{.5cm} \rho_{\pm}=a_{\pm}(T) T_{\pm}^4+\epsilon_{\pm}(T)\\
\end{equation}
where $p_{\pm}$ and $\rho_{\pm}$ are the pressure and energy density of the fluid in either phase, and
\begin{equation}\label{eq:eos}
\begin{aligned}
 a_{\pm}(T)\equiv -\frac{3}{4T^3}\frac{d V_{\rm eff}\left[\phi_{\pm}(T),T\right]}{dT}, \hspace{0.2cm} \epsilon_{\pm}(T) \equiv V_{\rm eff}\left[\phi_{\pm}(T),T\right]+\frac{1}{3} a_{\pm}(T)T^4.
\end{aligned}
\end{equation}
The above form for the equations, taken from Ref.~\cite{Espinosa:2010hh}, are inspired by the so-called `Bag EoS', but involves the temperature-dependent quantities $a_{\pm}(T)$, $\epsilon_{\pm}(T)$.  Fortunately, we can safely neglect the temperature dependence in $a_{\pm}$, $\epsilon_{\pm}$, using their values at $T=T_n$.  This is because the free energy (and hence $a_{\pm}$, $\epsilon_{\pm}$) are dominated by light degrees of freedom, which contribute a constant term to the free energy in each phase that does not vary significantly between $T_n$ and $T_c$ for the cases we consider. We find that using $a_{\pm}(T_n)$, $\epsilon_{\pm}(T_n)$ in Eq.~\ref{eq:eos} reproduces the full result for the pressure and energy density within a few percent.  This is fortunate as it allows us to avoid several issues arising for more complicated temperature dependence in the EoS, such as the variation of the sound speed in the plasma \cite{Leitao:2014pda}.

Ultimately we will take $T_{eq}(x)\equiv T_++\delta T_{bg}(x)$ to be the (space-time--dependent) temperature entering the equilibrium distribution functions for the various particles in the plasma.  We thus need to determine $T_+$, the temperature just outside the bubble. The pressure and energy density can be related to the fluid velocities on either side of the phase boundary by integrating Eq.~\ref{eq:fluid} across the wall.  This yields expressions for the fluid velocities $v_{\pm}$ in the wall frame which depend on $T_{\pm}$:
\begin{equation} \label{eq:vpm}
v_{+}v_-=\frac{p_+-p_-}{\rho_+-\rho_-}, \hspace{.5cm} \frac{v_+}{v_-}=\frac{\rho_-+p_+}{\rho_++p_-}.
\end{equation}
These velocities can be simply transformed to their analogs in the fluid frame $\widetilde{v}_{\pm}$ via $\widetilde{v}_{\pm}=v_w-|v_{\pm}|/(1-v_w |v_{\pm}|)$.  Note that for a subsonic deflagration, $\widetilde{v}_-=0$ and so $v_w=-v_-$, as discussed in e.g. Refs.~\cite{Enqvist:1991xw, Espinosa:2010hh}.  

One then needs to relate the temperature $T_{+}$ to $T_n$.  In the subsonic deflagration case, the bubble wall is preceded by a shock front moving with velocity $v_{sh}$ in the fluid frame. The temperatures $T_{1,2}$ and fluid velocities $v_{1,2}$ on either side of the shock front (in its rest frame) will be different; the equation of state is however the same (again neglecting small temperature variations in $a_+(T)$, $\epsilon_+(T)$).  We will use the subscripts 1,2 to denote quantities inside and outside the shock front, respectively.  One can again integrate across the interface and use the fact that the fluid is at rest beyond the shock front (i.e. $\widetilde{v}_2=0\rightarrow v_2=-v_{sh}$) with temperature $T_2=T_n$. This yields an expression for $v_1$ in terms of $T_1$ and $T_n$:
\begin{equation}
v_1^2=\frac{3T_n^4+T_1^4}{9T_1^4+3T_n^4}.
\end{equation}
Again, the corresponding fluid velocity in the fluid frame $\widetilde{v}_1$ is simply given by velocity addition, $\widetilde{v}_1=\frac{3v_1^2-1}{2v_1}$. 

Throughout our calculation we neglect the curvature of the bubble wall.  In this approximation, the temperatures and fluid velocities (in the fluid frame) between the bubble wall and the shock wave are simply constant \cite{Leitao:2010yw}, and so one can set 
\begin{equation}
\widetilde{v}_1(T_1,T_n)\approx \widetilde{v}_{+}(T_+,v_w), \hspace{0.3cm} T_1\approx T_+ 
\end{equation}
and solve for $T_+$ in terms of $T_n$, $v_w$.  
Previous studies suggest that using the planar approximation instead of the full solutions to the spherical hydrodynamic equations can reproduce the full result for the wall velocity to within a few percent~\cite{Huber:2013kj}.

With the temperature $T_+$ and the static properties of the phase transition determined in this way, we can now consider the asymptotic behavior of the bubble after its formation.
 
\subsection{Wall Equations of Motion}

 \begin{figure}[!t]
\centering
\includegraphics[width=.4\textwidth]{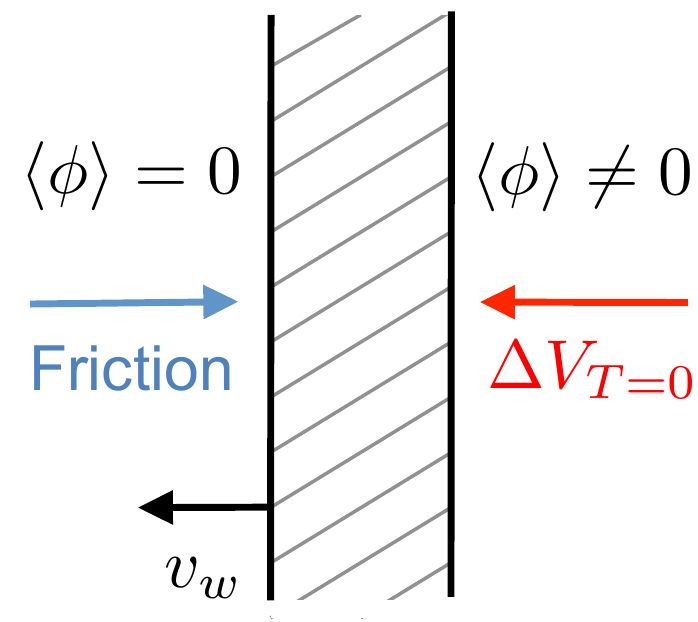}
\caption{\label{fig:forces} Illustration of the competing forces acting on the bubble wall that ultimately determine $v_w$. The steady state wall velocity is such that the vacuum energy difference between the phases ($\Delta V_{T=0}$) is balanced by the friction provided by the interactions of the wall with the plasma.}
\end{figure}

The main object for our analysis will be the bubble wall equations of motion corresponding to the set of scalar fields $\phi_i=\phi_h,\phi_s$.  These can be derived by requiring conservation of the energy-momentum tensor for the scalar field condensates computed in a WKB approximation \cite{Moore:1995si}, or directly from the Kadanoff-Baym equations \cite{Konstandin:2014zta}. We are interested in the stationary limit of the equations of motion in the plasma frame; that is, we want to investigate the bubble wall once it has reached its terminal velocity (if it exists), with the pressure driving the expansion precisely counterbalanced by the drag force exerted on the bubble by the plasma. This is illustrated in Fig.~\ref{fig:forces}.

Neglecting the curvature of the bubble, in the rest frame of a stationary (non-accelerating) bubble wall all functions will be depend only on $z$, the distance from the phase boundary.  Consequently, in the plasma frame, all functions depend only on the coordinate $x\equiv z+v_w t$, where $v_w$ is the wall velocity in the plasma frame and we have assumed that the wall is moving to the left.  In the stationary wall limit, the equations of motion then simplify to 
\begin{equation} \label{eq:wall_eom_1}
\begin{aligned}
-(1&-v_w^2)\phi_i'' + \frac{\partial V(\phi_i,T)}{\partial \phi_i} +\sum_j \frac{\partial m_j^2(\phi_i)}{\partial\phi_i} \int \frac{d^3 p}{(2\pi)^3 2 E_j} \delta f_j(p,x) =0
\end{aligned}
\end{equation}
where primes indicate differentiation with respect to $x$.  Here the sum is over all fields coupling to the scalar field $\phi_i$, $E_j$ is the (space-time--dependent) energy of the particle $j$, $E_j=\sqrt{p^2+m_j^2(x)}$, and $\delta f_j$ is the deviation from the equilibrium distribution function for the species $j$.  

Solutions to the above equations of motion typically only exist for one subsonic value of the constant $v_w$. This is the quantity we wish to determine.  To do so, one must find profiles $\phi_i(x)$ such that Eq.~\ref{eq:wall_eom_1} is satisfied, which in turn requires solving for the deviations from equilibrium of the various species in the plasma. These deviations, along with the equilibrium contributions, are responsible for the drag force on the bubble wall. Unfortunately, the $\delta f_j$ depend non-trivially on $v_w$ and the bubble profile, so Eq.~\ref{eq:wall_eom_1} represents a set of integro-differential equations. 

\subsection{Aside: Runaway Bubbles and Tree-level Cubic Terms}

Before moving on to the case of non-relativistic bubbles (relevant for electroweak baryogenesis), we can begin by considering the wall dynamics in a simple limit: that of ultra-relativistic, ``runaway'' bubbles \cite{Bodeker:2009qy}, with Lorentz factor $\gamma \gg 1$. In this case, the friction on the bubble from the plasma in the large-$\gamma$ limit is too small to counterbalance the pressure difference between the vacua, which drives the expansion. Ref.~\cite{Bodeker:2009qy} showed that this situation is common in singlet-driven transitions, so it is important to review this case before moving on to the non-relativistic regime.

Following Ref.~\cite{Bodeker:2009qy}, a runaway solution to the equations of motion exists provided
\begin{equation}\label{eq:runaway}
V_{\rm eff}(T=0,\phi_{+})-V_{\rm eff}(T=0,\phi_-)+\sum_i N_i \left[ m_i^2(\phi_+)-m_i^2(\phi_-)\right] \int \frac{d^3 p}{(2\pi)^3 2 E}f_{0,i}(p,\phi_+)>0
\end{equation}
at the nucleation temperature. Here, $f_0$ is the equilibrium distribution function of the species $i$, and $\phi_{\pm}$ are the field values at the minima of the potential. In the high-$T$ limit, there is a simple interpretation of this criterion in terms of the high-temperature expansion of the thermal effective potential: a runaway solution will exist if it is energetically favorable to tunnel to the broken phase in the `mean-field' potential, obtained by retaining only the $T^2$ terms in Eq.~\ref{eq:pot}. In other words,
\begin{equation}\label{eq:runaway_cond}
V_{\rm eff}^{\rm no \hspace{0.1cm} cubic}(\phi_+,T_n)>V_{\rm eff}^{\rm no \hspace{0.1cm} cubic}(\phi_-,T_n) \hspace{0.3cm} \Rightarrow \hspace{0.3cm} {\rm runaway \hspace{.1cm}solution \hspace{0.1cm} exists.}
\end{equation}

The above expression indicates that all points found with a first-order phase transition in our gauge-invariant approach (retaining only the quadratic finite-$T$ terms) would feature an ultra-relativistic wall solution if there were no other contributions to the effective potential. This may appear incompatible with our goal of determining subsonic solutions to the equations of motion but it is not. First of all, including the finite temperature cubic term inevitably changes the the transition temperature and the effective potential at that temperature. This can cause the same parameter space point to instead feature $V_{\rm eff}^{\rm no \hspace{0.1cm} cubic}(\phi_+,T_n)<V_{\rm eff}^{\rm no \hspace{0.1cm} cubic}(\phi_-,T_n)$ , and hence no runaway solution. We indeed find this to be the case for most points considered when including the gauge boson cubic term in our parameter scans. Even if a runaway solution exists for the EOMs including the full finite-$T$ effective potential, there is another important caveat. The criterion in Eq.~\ref{eq:runaway_cond} assumes that the bubble is in the ultra-relativistic regime to begin with. However it is instead possible for the friction to prevent the bubble from ever reaching such large velocities required for Eq.~\ref{eq:runaway} to be valid. In fact, hydrodynamic effects alone can obstruct the wall from expanding ultra-relativistically \cite{Konstandin:2010dm}. Thus, even if a particular parameter space point admits a runaway solution, it may not be realized if a subsonic stationary solution exists.  On the other hand, even if no runaway solution exists, one with $v_w>c_s$ might. The reader should thus bear in mind that our approach will find subsonic solutions to the equations of motion, not guarantee that they are realized. This is also true of previous studies~\cite{Moore:1995ua,  Moore:1995si, John:2000zq, Konstandin:2014zta}.

Equation~\ref{eq:runaway} shows that the friction force acting on the wall takes a very simple form in the $\gamma \gg 1$ limit. This is not the case for the subsonic walls we are interested in. Determining the wall velocity in the $\gamma\sim 1$ regime requires a careful calculation of the various deviations from equilibrium in the plasma. This is what we discuss in the following section. 

\section{Kinetic Theory and Deviations from Equilibrium}\label{sec:Boltz}

\subsection{Setup}
With the temperature $T\equiv T_+$ inferred from hydronamic considerations, the first step towards solving the bubble wall equations of motion in the non-relativistic ($\gamma \approx 1$) case is determining the distribution functions $f_i$ for the various excitations appearing in Eq.~\ref{eq:wall_eom_1}.  To do so, we will primarily utilize a perturbative effective kinetic theory approach \cite{Jeon:1995zm, Arnold:2002zm}, as in previous studies \cite{Moore:1995ua, Moore:1995si, John:2000zq} (we will take a somewhat different approach for the corresponding gauge boson friction, which should be modeled classically as discussed below). This treatment applies to weakly coupled excitations with local interactions and short wavelengths compared to the length scale of the bubble wall in the plasma frame, i.e.
\begin{equation} \label{eq:WKB}
E \gg \frac{1}{L_w}
\end{equation}
where $L_w$ is the wall width.  Typical momenta are of order $p\sim T$, but softer excitations will be present in the plasma as well. We will assume that the kinetic theory description is viable in the range $p\gtrsim g T$, which is reasonable for the particles we will be interested in given the values we find for the wall widths. Here and throughout this section $g$ represents a generic dimensionless coupling of the theory\footnote{The coupling $g$ should be thought of as some combination of couplings entering the thermal and zero temperature masses of the particle in question. In other words, we assume parametrically that $m\sim g T \sim g \phi$ near the electroweak phase transition.} that is assumed to be small. Infrared (IR) excitations with momenta $p \ll T$ will not be captured by this treatment, since their interactions cannot be properly described by a local collision term. These contributions can be important for the bosonic species \cite{Moore:2000wx}, but the perturbative effective kinetic theory should provide an adequate estimate of the damping force on the bubble wall, provided that very infrared excitations are equilibrated quickly  \cite{Moore:1995si}, as we will assume for most of the species we are interested in\footnote{This was shown to be a poor assumption for the $SU(2)_L$ gauge bosons in Ref.~\cite{Moore:2000wx}, which we discuss further in Sec.~\ref{sec:gauge}. Infrared contributions from the Higgs and singlet fields may be important. However, their equations of motion are not over-damped as they are for the gauge bosons \cite{Moore:2000wx}, and so their distributions should equilibrate more quickly than those for the gauge fields.}. 

In the effective kinetic theory we consider, the quasiparticle distribution function for the species $i$ satisfies the Boltzmann equation
\begin{equation} \label{eq:dtBoltz}
\frac{d}{dt}f_i \equiv \left(\frac{\partial}{\partial t}+\dot{z}\frac{\partial}{\partial z}+\dot{p_z}\frac{\partial}{\partial p_z} \right)f_i = -C[f]_i
\end{equation}
in the fluid frame, where $C[f]_i$ is a local collision integral. The collision term involves all interactions of the species $i$ with all other excitations in the plasma. It can be written as \cite{Arnold:2002zm}
\begin{equation}\label{eq:coll}
\begin{aligned}
C[f]_i=\frac{1}{2 N_i} \sum_{jmn} \frac{1}{2E_p}\int \frac{d^3k d^3 p^{\prime} d^3 k^{\prime}}{(2\pi)^9 2E_k 2E_{p^{\prime}} 2E_{k^{\prime}}} &\left| \mathcal{M}_{ij\rightarrow mn} (p,k;p^{\prime},k^{\prime})\right|^2 (2\pi)^4 \delta(p+k-p^{\prime}-k^{\prime}) \\ 
&\times \mathcal{P}_{ij\rightarrow m n} \hspace{0.1cm} [f_i(p),f_j(k),f_m(p^{\prime}),f_n(k^{\prime})]
\end{aligned}
\end{equation}
where the sum is over all 4-body processes $ij\rightarrow mn$, with the momenta labeled as $p$, $p^{\prime}$, $k^{\prime}$, and $k$ moving clockwise around the diagram starting with particle $i$. The matrix elements include finite-temperature effects (discussed below) and are summed over helicities and colors of all four external quasiparticles, then divided by the number of degrees of freedom corresponding to species $i$, $N_i$ ($N_h=1$, $N_t=N_{\bar{t}}=6$)~\footnote{We will neglect any possible $CP$-violation coupling to the top quark and hence assume that the top and anti-top densities are identical. This means we can compute the top perturbations and simply count their contribution to the condensate equations of motion twice.}. The population factor is
\begin{equation}
 \mathcal{P}_{ij\rightarrow m n} \equiv f_i f_j(1\pm f_m)(1\pm f_n)-f_m f_n(1\pm f_i)(1\pm f_j)
\end{equation}
with the upper (lower) signs corresponding to bosons (fermions) and $f_a$ the appropriate Bose-Einstein or Fermi-Dirac distribution function for particle $a$, which we assume to take the form
\begin{equation}
f_a=\left(e^{(E+\delta_a)/T}\pm 1\right)^{-1}.
\end{equation}
In Eq.~\ref{eq:coll}, the prefactor of $1/2$ takes care of both the symmetry factor when identical particles are present in the final state, and the double counting that occurs from the unrestricted sum over $m$ and $n$.

The Boltzmann equations above apply to all quasiparticles in the plasma satisfying Eq.~\ref{eq:WKB} with sufficiently high momentum. However, examining Eq.~\ref{eq:wall_eom_1}, we see that only the distribution functions of field excitations with significant couplings to the relevant scalar fields involved in the phase transition are required. Since these particles have significant couplings to the Higgs and singlet scalar fields, we will refer to them as `heavy'. Also, $\delta f_i=\delta f_i(p,x)$ has some space-time--dependence, arising in part from the spatial variation of the background fluid temperature and velocity across the bubble wall, as discussed in Sec.~\ref{sec:hydro}. The background fluid is in \emph{local} thermal equilibrium and comprises all `light' effective degrees of freedom. Note that quasiparticles with large field-independent masses will be irrelevant for our purposes, since their distribution functions feature significant Boltzmann suppression. Also, precisely which fields should be considered `heavy', `light', or irrelevant depends on the given model. For the singlet-driven scenarios we are concerned with here, the heavy fields will be the top quarks, gauge, Higgs, and singlet bosons.

To find approximate solutions to the Boltzmann equations for the heavy species and background, we will utilize the \emph{`fluid ansatz'}~\cite{Moore:1995si}, in which case the perturbations are assumed to take the form
\begin{equation}
\delta_j = - \mu_{j}-\frac{E}{T}(\delta T_j+\delta T_{\rm bg})-p_z(\delta v_j +  v_{\rm bg}).
\end{equation}
Here $ \mu_j$, $\delta T_j$, $\delta v_j$ are the chemical potential, temperature perturbation, and velocity perturbation of the species $j$, respectively, in the plasma frame.  We have assumed that the fields with small couplings to the scalar condensates $\phi_{h,s}$ are in thermal equilibrium at a common space-time--dependent temperature $T_++\delta T_{\rm bg}(x)$ and velocity $v_{\rm bg}(x)$ with vanishing chemical potential, as in Ref.~\cite{Moore:1995si}. The assumption that $\mu_{\rm bg}\approx 0$ is valid whenever the total background particle destruction rate is larger than that for the heavy particles, as will be the case here (all pure gluon rates are enhanced by the large color factors and Bose statistics). The space-time--dependence in $\delta T_{\rm bg}$, $v_{\rm bg}$ arises from the change in masses of the corresponding particles moving from the $\phi_i\neq 0$ phase inside the bubble to the $ \phi_i = 0$ vacuum outside. 

Throughout this study, we will work to linear order in the perturbations, which are assumed to be small ($\mu_j/T$, $\delta T_j/T$, $\delta T_{\rm bg}/T$, $\delta v_j$, $v_{\rm bg} \ll1$). This should be the case for moderately strong phase transitions, and we verify the validity of this assumption \textit{a posteriori}. It should be noted that this treatment can be extended to accommodate large fluid velocities in front of the wall~\cite{Konstandin:2014zta}, although this will not be necessary for any of the transitions we consider. As a result, we set all Lorentz $\gamma$ factors to 1 throughout our calculation.    

With the above definitions, the population factor $\mathcal{P}$ is given to linear order in the perturbations by
\begin{equation}
\mathcal{P}\simeq f_{0,1} f_{0,2} (1\pm f_{0,3})(1\pm f_{0,4})  \left( \delta_1 + \delta_2 - \delta_3 -\delta_4\right)
\end{equation}
where the `0' subscript indicates the corresponding equilibrium distribution function. Note that the background temperature and velocity perturbations do not enter the collision integrals to linear order.

To determine $\mu_i$, $\delta T_i$, and $\delta v_i$ we follow Refs.~\cite{Moore:1995ua, Moore:1995si, John:2000zq} and take three moments of each equation, multiplying by $\int d^3p/(2\pi)^3$,  $\int E_p/T d^3p/(2\pi)^3$,  $\int p_z/T d^3p/(2\pi)^3$ and solve the resulting expressions for the perturbations. For a given heavy species, the relevant three equations are given in the plasma frame by
\begin{equation}\label{eq:Boltz}
\begin{aligned}
c_2^i \frac{\partial}{\partial t} \mu_i +c_3^i \frac{\partial}{\partial t}(\delta T_i + \delta T_{\rm bg})+\frac{c_3^i T}{3}\frac{\partial}{\partial z}(\delta v_i + v_{\rm bg})+\int \frac{d^3 p}{(2\pi)^3 T^2}C[f]_i &=\frac{c_1^i}{2T}\frac{\partial m^2_i}{\partial t}\\
c_3^i \frac{\partial}{\partial t} \mu_i +c_4^i \frac{\partial}{\partial t}(\delta T_i + \delta T_{\rm bg})+\frac{c_4^i T}{3}\frac{\partial}{\partial z}(\delta v_i + v_{\rm bg})+\int \frac{E d^3 p}{(2\pi)^3 T^3}C[f]_i &=\frac{c_2^i}{2T}\frac{\partial m^2_i}{\partial t}\\
\frac{c_3^i}{3} \frac{\partial}{\partial z} \mu_i +\frac{c_4^i}{3} \frac{\partial}{\partial z}(\delta T_i + \delta T_{\rm bg})+\frac{c_4^i T}{3}\frac{\partial}{\partial t}(\delta v_i + v_{\rm bg})+\int \frac{p_z d^3 p}{(2\pi)^3 T^3}C[f]_i &=0
\end{aligned}
\end{equation}
where an ingoing particle of the relevant species has momentum $p$ and where
\begin{equation}
c_n^i \equiv \int \frac{E^{n-2}}{T^{n+1}}(-f_{0,i}')\frac{d^3 p}{(2\pi)^3}.
\end{equation}  Further details can be found in Ref.~\cite{Moore:1995si}.  The resulting collision terms for each heavy field $i$ can be written as 
\begin{equation}
\begin{aligned}
\int \frac{d^3 p}{(2\pi)^3 T^2} C[f]_i \equiv &\sum_j \left(\delta \mu_j \Gamma_{\mu_1,j}^i+\delta T_i \Gamma_{T_1,j}^i\right)\\
\int \frac{d^3 p}{(2\pi)^3 T^3}E_i C[f]_i \equiv &\sum_j\left( \delta \mu_j \Gamma_{\mu_2,j}^i+\delta T_i \Gamma_{T_2,j}^i\right)\\
\int \frac{d^3 p}{(2\pi)^3 T^4} p_{z,i} C[f]_i \equiv &\sum_j \left(\delta v_j \Gamma_{v_1,j}^i\right)
\end{aligned}
\end{equation}

The background excitations also satisfy a set of Boltzmann equations, 
\begin{equation}\label{eq:bkg_Boltz}
\begin{aligned}
\sum c_4 \left( \frac{\partial}{\partial t}\delta T_{\rm bg}+\frac{c_4 T}{3}\frac{\partial}{\partial z} v_{\rm bg}\right)+\int \frac{E d^3 p}{(2\pi)^3 T^3}C[f]_{\rm bg} &=0\\
\sum \frac{c_4}{3} \left(\frac{\partial}{\partial z} \delta T_{\rm bg}+T\frac{\partial}{\partial t} v_{\rm bg}\right)+\int \frac{p_z d^3 p}{(2\pi)^3 T^3}C[f]_{\rm bg} &=0
\end{aligned}
\end{equation}
which arise from Eq.~\ref{eq:Boltz} with $\mu_{\rm bg}\approx 0$. The sum above is over all background species, with $\overline{c}_4\equiv \sum c_4$ the heat capacity of the plasma. As for the heavy quasiparticles, the collision terms can be written as
\begin{equation}\label{eq:bkg_int}
\begin{aligned}
\int \frac{d^3 p}{(2\pi)^3 T^3}E_i C[f]_{\rm bg} \equiv &-\sum_j\left( \delta \mu_j \widetilde{\Gamma}_{\mu_2,j}+\delta T_i \widetilde{\Gamma}_{T_2,j}\right)\\
\int \frac{d^3 p}{(2\pi)^3 T^4} p_{z,i} C[f]_{\rm bg} \equiv &-\sum_j \left(\delta v_j \widetilde{\Gamma}_{v_1,j}\right)
\end{aligned}
\end{equation}
Although $\delta T_{\rm bg}$ and $v_{\rm bg}$ do not enter the collision integrals, the perturbations corresponding to the heavy excitations do. The convention for evaluating the matrix elements is the same as for the heavy particles, with all background excitations treated as one species. Thus, every heavy particle process involving the background excitations will contribute to Eqs.~\ref{eq:bkg_int}. We will calculate all of the contributions relevant for singlet-driven transitions in the next subsection.

\subsection{Relevant Excitations and Interaction Rates}

In the SM and its singlet extensions, the relevant heavy species to consider in Eq.~\ref{eq:wall_eom_1} above are typically the top quarks,  $SU(2)_L$ gauge, Higgs, and singlet bosons, with the Higgs and singlet excitations being the dominant source of friction on the singlet field condensate.  We will consider two different sets of contributions to the total friction. For the gauge-invariant calculation, we include only the top quark, Higgs, and singlet contributions. When incorporating the gauge boson cubic term, we will also account for the friction arising from the $SU(2)_L$ gauge bosons. We will not include the Goldstone friction contribution, since we drop the corresponding finite-$T$ cubic term from the effective potential. This should be a reasonable approximation as the Higgs excitations will only make up about $20\%$ of the total friction on the wall\footnote{We do include the Goldstones, $\phi^0$, $\phi^{\pm}$ in the various interaction rates. They only appear as external legs of the diagrams we consider, and the corresponding matrix elements are gauge-independent. They are treated as a background species.}. 

The friction from each species enters the bubble wall EOM (Eq.~\ref{eq:wall_eom_1}) through the derivative of the corresponding mass squared. For the top quarks, the effective mass squared is 
\begin{equation}
m_t^2(\phi_h)=\frac{1}{2} y_t^2 \phi_h^2+\Pi_t(T)\\
\end{equation}
with the corresponding thermal self-energy correction
\begin{equation}
\Pi_t(T)\simeq \frac{1}{6} g_3^2 T^2\\,
\end{equation}
neglecting the subdominant thermal $SU(2)_L$ and $U(1)_Y$ contributions.  

The Higgs and singlet require slightly more care. Throughout the remainder of this study we will neglect all mixing effects between the SM-like Higgs and singlet excitations. As discussed in Sec.~\ref{sec:pheno}, we will choose the parameters of the $T=0$ Lagrangian such that the mixing vanishes in the broken electroweak phase. At finite temperature and across the bubble wall this will no longer be the case. However, in the high temperature limit, the effective neutral scalar mass matrix is diagonal, since off-diagonal thermal corrections are proportional to dimensionful parameters and vanish as $T\rightarrow \infty$. For temperatures around the electroweak phase transition, the thermal masses still dominate the mixing matrix, and so this should be a decent approximation. The relevant field-dependent masses, including the leading thermal corrections, are then
\begin{equation}\label{eq:masses}
\begin{aligned}
&m_h^2(\phi_h,\phi_s)\simeq-\mu^2+3\lambda \phi_h^2+\frac{1}{2}a_1 \phi_s +\frac{1}{2} a_2 \phi_s^2+\Pi_h(T)\\
&m_s^2(\phi_h,\phi_s)\simeq b_2+2 b_3 \phi_s + 3 b_4 \phi_s^2 +\frac{1}{2}a_2 \phi_h^2+\Pi_s(T)
\end{aligned}
\end{equation}
with the thermal masses
\begin{equation}\label{eq:therm_mass}
\begin{aligned}
&\Pi_h(T)\simeq \left(\frac{3}{16}g_2^2+\frac{1}{16} g_1^2+\frac{1}{4}y_t^2+\frac{1}{2}\lambda+\frac{1}{24}a_2\right)T^2\\
&\Pi_s(T)=\left(\frac{1}{6}a_2+\frac{1}{4}b_4\right)T^2
\end{aligned}
\end{equation}
where we have left out the light fermion Yukawa contributions. Since we are neglecting the finite-temperature tadpole contribution to the effective potential we also drop the $a_1$ terms in the masses above. This is required for consistency, since these terms are precisely those that give rise to the finite-temperature tadpole.  Finally, as in Ref.~\cite{Moore:1995si}, we treat the transverse $SU(2)_L$ gauge bosons as a single species $W$ with field-dependent mass squared
\begin{equation}
m_W^2(\phi_h) = \frac{1}{4} g_2^2 \phi_h^2.
\end{equation}
Transverse excitations do not acquire a thermal mass at leading order in the couplings. Longitudinal modes obtain an effective thermal (Debye) mass at leading order, corresponding to the inverse screening length of electric potentials in the plasma~\cite{Comelli:1996vm}. This is given by
 \begin{equation}
 m_{D,W}^2(T)\simeq \frac{11}{6}g_2^2 T^2
 \end{equation}
 in the Standard Model. Since the gauge boson friction is dominated by very infrared excitations, only the transverse contributions will be relevant.
 
Our strategies for dealing with each of these types of excitations will differ. As we will see below, the top quark and Higgs interaction rates are typically sizable, and so the collision term plays an important role in the corresponding Boltzmann equations. This is not expected to be the case for singlet quasiparticles at high temperature. Contrary to the tops and Higgs, we will assume that the singlet interactions are slow. In this case, the collision term can be neglected. The corresponding Boltzmann equation decouples from the rest of the system and can be solved exactly. We discuss this further in Secs.~\ref{sec:singlet} and \ref{sec:singlet_sol}. Finally, the gauge boson contributions are dominated by infrared dynamics and require a classical treatment, which has been worked out in Ref.~\cite{Moore:1995si} and discussed in Sec.~\ref{sec:gauge} below.

Let us first consider the interactions involving the top quark, Higgs, and background excitations.   

\subsubsection{Top, Higgs, and Background Excitations}

Solving the Boltzmann equations for the perturbations $\mu_{t,h}$, $\delta T_{t,h}$, $\delta T_{\rm bg}$, $\delta v_{t,h}$, and $v_{\rm bg}$ requires computing the collision integrals corresponding to all the four-body interactions involving $t$, $h$, and the background fields. This task is rather daunting due to the sheer number of allowed processes. However, the dominant interactions will be of $\mathcal{O}(\alpha_s^2)$ for the top quarks, and $\mathcal{O}(\alpha_s \alpha_t)$, $\mathcal{O}(\alpha_t^2)$ for the Higgs bosons, where $\alpha_s=g_3^2/4\pi$, $\alpha_t=y_t^2/4\pi$. We will therefore focus on these interactions, neglecting, for example, contributions involving a factor of $\alpha_w$, which are numerically small compared to the Yukawa-type contributions for the Higgs bosons\footnote{We have verified this is the case despite the enhancement provided by Bose-Einstein statistics.}. 

To estimate the relevant interaction rates, we will work at leading order in all couplings in the high-$T$, weak coupling limit, neglecting all terms of $\mathcal{O}(m^2/T^2)$ (here $m$ should be understood as either a zero-temperature or thermal mass). This is the approximation used in all previous microphysical studies of the wall velocity \cite{Moore:1995ua, Moore:1995si, John:2000zq}, as well as in the context of plasma properties in arbitrary high-temperature gauge theories \cite{Arnold:2000dr, Arnold:2003zc}. This approximation can begin to break down inside the bubble wall for the top quarks and scalars and could in principle be improved upon in the future. Nevertheless, it should reproduce at least the correct parametric dependence of the full leading order result. In this limit we can neglect the effect of the space-time--varying masses on the interaction rates.  Hard excitations dominate the phase space  for the relevant top quark and Higgs collision integrals, and can be characterized by massless dispersion relations
\begin{equation}
E^{\rm hard}=\sqrt{p^2+m^2+m_{\rm th}^2(T)}\simeq p+\mathcal{O}\left(gT\right)
\end{equation}
to leading order in the couplings and in the high-temperature limit.

Although the external quasiparticles can be treated as massless in this approximation, infrared excitations appearing as mediators in $t$- and $u$-channel diagrams naively result in logarithmic IR divergences in the Higgs and top quark scattering amplitudes.  These divergences are cut off by the interactions of the mediator with the plasma. For long-wavelength excitations, the corrections comprise so-called `hard thermal loops' (HTLs), and result in a breakdown of the perturbative expansion. The corrections can be resummed into a thermal self-energy correction to the propagator, valid in the low momentum limit\footnote{Note that the same situation arises when computing loop corrections to the finite-temperature effective potential.}. The self-energy is typically of order $g T$ and so these processes can produce sizable logarithmic enhancements of the corresponding matrix elements, scaling as $\sim g(T)^4 \log1/g(T)$ at high temperatures.  This provides a useful way of categorizing the most important diagrams contributing to $\mathcal{M}_i$ in $C[f]$ in the high-$T$ limit.

A full leading-order determination of the effective scattering rates in the plasma is possible \cite{Jeon:1994if, Arnold:2003zc}, though computationally more involved and beyond the scope of this work. It would be interesting to revisit in the future. We will instead work in a `leading logarithm' expansion, keeping only contributions of order $\sim g(T)^4 \log1/g(T)$, which are typically the largest. In this approximation, only 4-body rates with $t$- and $u$-channel diagrams contribute. For further details on this approximation, see Refs.~\cite{Moore:1995si, Arnold:2000dr}.

Another subtlety arises in computing scattering rates in the high-$T$ limit involving soft $t$- or $u$-channel exchange. The thermal self-energies involved in the propagators are generally momentum-dependent \cite{Arnold:2003zc}. Previous studies of the wall velocity neglected these contributions,  simply replacing them with the corresponding Debye (thermal) masses for the corresponding gauge bosons (fermions).  However, including the momentum-dependent self-energies, which enter at leading order in the couplings, can have a significant effect and has been shown to often provide better agreement between the leading log and full leading order results for plasma transport coefficients in high-temperature gauge theories \cite{Arnold:2003zc}. Consequently, we will use the full momentum-dependent HTL-resummed propagators \cite{Klimov:1981ka, Weldon:1982bn, Arnold:2003zc} in computing the various collision integrals for the top quark and Higgs excitations.

 \begin{table}[!t]
\centering
 \begin{tabular}{l c c }
 Process &$\left|\mathcal{M}\right|_{\rm tot}^2$ & Internal Propagator\\
 \hline
 \hline

\vspace{0.2cm}  \uline{$\mathcal{O}(g_3^4)$}:&&\\

\vspace{0.2cm} $t\bar{t} \leftrightarrow gg$: \hspace{0.3cm} &$\frac{128}{3}  g_3^4 \left( \frac{u}{t}+\frac{t}{u}\right) $& $t$\\

\vspace{0.2cm} $tg \leftrightarrow tg$: \hspace{0.3cm} &-$\frac{128}{3}  g_3^4 \frac{s}{u}+96 g_3^4 \frac{s^2+u^2}{t^2}$& $g,t$\\

\vspace{0.2cm} $tq (\bar{q}) \leftrightarrow tq (\bar{q})$: \hspace{0.3cm} &$160  g_3^4 \frac{u^2+s^2}{t^2}$ & $g$\\

\vspace{0.2cm}  \uline{$\mathcal{O}(y_t^2 g_3^2)$}:&&\\

\vspace{0.2cm} $t\bar{t} \leftrightarrow hg$, $\phi^0 g$: \hspace{0.3cm} &$8 y_t^2 g_3^2 \left(\frac{u}{t} +\frac{t}{u}\right)$& $t$\\

\vspace{0.2cm} $t\bar{b} \leftrightarrow \phi^+g:$&$8 y_t^2 g_3^2 \left(\frac{u}{t} +\frac{t}{u}\right)$ & $t$, $b$\\

\vspace{0.2cm} $t g \leftrightarrow th$, $t \phi^0$:&$-8 y_t^2 g_3^2 \frac{s}{t}$ &$t$ \\

\vspace{0.2cm} $t g \leftrightarrow b \phi^+$:&$-8 y_t^2 g_3^2 \frac{s}{t}$ &$b$ \\

\vspace{0.2cm} $t \phi^-\leftrightarrow bg$: &$-8 y_t^2 g_3^2 \frac{s}{t} $ &$t$\\

\vspace{0.2cm}  \uline{$\mathcal{O}(y_t^4)$}:&&\\

\vspace{0.2cm} $t\bar{t} \leftrightarrow hh$, $\phi^0 \phi^0$: \hspace{0.3cm} & $\frac{3}{2} y_t^4\left( \frac{u}{t} + \frac{t}{u}\right)$& $t$\\

\vspace{0.2cm} $t\bar{t} \leftrightarrow \phi^+ \phi^-$: \hspace{0.3cm} &$3 y_t^4 \frac{u}{t} $& $b$\\

\vspace{0.2cm} $t\bar{t} \leftrightarrow h \phi^0$: \hspace{0.3cm} &$\frac{3}{2} y_t^4\left( \frac{u}{t} + \frac{t}{u}\right)$ & $t$\\

\vspace{0.2cm} $t\bar{b} \leftrightarrow h \phi^+$, $\phi^0 \phi^+$: \hspace{0.3cm} &$\frac{3}{2} y_t^4 \frac{u}{t} $& $t$\\

\vspace{0.2cm} $th$, $t \phi^0 \leftrightarrow h t$, $\phi^0 t$: \hspace{0.3cm} &$-\frac{3}{2} y_t^4 \frac{s}{t} $& $t$\\

\vspace{0.2cm} $t\phi^- \leftrightarrow h b$, $\phi^0 b$: \hspace{0.3cm} &$\frac{3}{2} y_t^4 \frac{u}{t} $& $t$\\

\vspace{0.2cm} $t\phi^+ \leftrightarrow \phi^+ t$: \hspace{0.3cm} &$3 y_t^4 \frac{u}{t} $& $b$\\

\end{tabular}

\caption{Relevant 4-body processes and their corresponding matrix elements in the leading log approximation.  The matrix elements are summed over the helicities and colors of all four external states (as well as flavors and quark -- anti-quark for $t q \rightarrow t q$). The excitation appearing on the internal propagators in each case is listed in the right-hand column. Note that other $t$- and $u$-channel processes exist, but do not contribute logarithmically to the collision integrals or are suppressed by powers of couplings small compared to $g_3$, $y_t$. }
\label{tab:m}
\end{table}

The relevant processes and their associated vacuum matrix elements are listed in Table~\ref{tab:m}. All terms of $\mathcal{O}(m^2/T^2)$ have been dropped. The leading log matrix elements are summed over the helicities and colors of the external particle (but not particle-antiparticle). These contributions will also be divided by the number of degrees of freedom of the species under consideration when entering the various $\Gamma^i_{k,j}$. 

The vacuum matrix elements must be modified to include the medium-dependent effects discussed above. At leading order, this amounts to inserting the momentum-dependent HTL self-energies on the internal lines. To translate the vacuum matrix elements above to their finite-temperature analogs, we can use the results of Refs.~\cite{Arnold:2002zm, Arnold:2003zc}.  For diagrams with an exchanged fermion in the leading log approximation, this amounts to the replacement 
\begin{equation}
\frac{u}{t} \simeq -\frac{s}{t} \rightarrow \frac{4\operatorname{Re}(p\cdot \widetilde{q} \hspace{0.1cm} k\cdot \widetilde{q}^*+s \widetilde{q}\cdot \widetilde{q}^*)}{\left|\widetilde{q}\cdot \widetilde{q}\right|^2}
\end{equation}
with $\widetilde{q}^{\mu}=p^{\mu}-p^{\prime \mu}-\Sigma ^{\mu}(p-p^{\prime})$ and $\Sigma^{\mu}(q)$ the fermionic HTL self-energy function \cite{Klimov:1981ka, Weldon:1982bn, Arnold:2003zc}
\begin{equation}
\begin{aligned}
&\Sigma^0(q)=\frac{m_f^2(T)}{2\left|\mathbf{q}\right|} \left(\log \frac{\left|\mathbf{q}\right|+q^0}{\left|\mathbf{q}\right|-q^0}-i \pi\right)\\
&\mathbf{\Sigma}(q)=\frac{-m_f^2(T) \hspace{0.1cm} \widehat{\mathbf{q}}}{\left|\mathbf{q}\right|}\left(1+i \pi -\frac{q^0}{2\left|\mathbf{q}\right|}\log \frac{\left|\mathbf{q}\right|+q^0}{\left|\mathbf{q}\right|-q^0} \right)
\end{aligned}
\end{equation}
with $m_f(T)$ the leading order fermion thermal mass, given approximately by $m_f(T)\approx \sqrt{1/6}g_3 T$ for quarks (see Eq.~\ref{eq:therm_mass}).  

For the gluon exchange diagrams, we must replace
\begin{equation}
\frac{s^2+u^2}{t^2} \rightarrow \frac{1}{2}\left(1+\left|D_{\mu \nu}(p-p^{\prime})(p+p^{\prime})^{\mu}(k+k^{\prime})^{\nu}\right|^2 \right)
\end{equation}
with the retarded thermal equilibrium gluon propagator $D_{\mu\nu}(q)$ given by
\begin{equation}
\begin{aligned}
&D_{00}(q)=\frac{-1}{\left|\mathbf{q}\right|^2+\Pi_{00}(q,T)}\\
&D_{ij}(q)=\frac{\delta_{ij}-\widehat{\mathbf{q}}_i\widehat{\mathbf{q}}_j}{q^2+\Pi_T(q,T)}\\
&D_{i0}(q)=D_{i0}(q)=0.
\end{aligned}
\end{equation}
The relevant HTL gauge boson self energy is \cite{Klimov:1981ka, Weldon:1982bn, Arnold:2003zc}
\begin{equation}
\begin{aligned}
&\Pi_{00}(q)=m_{\rm D}^2(T)\left(1-\frac{q^0}{2\left|\mathbf{q}\right|}\log \frac{\left|\mathbf{q}\right|+q^0}{\left|\mathbf{q}\right|-q^0}+\frac{i \pi q^0}{2\left|\mathbf{q}\right|}\right)\\
&\Pi_{T}(q)=m_{\rm D}^2(T)\left[\frac{q^0}{2\left|\mathbf{q}\right|}+\frac{q^0 q^2}{4\left|\mathbf{q}\right|^3}\left(\log \frac{\left|\mathbf{q}\right|+q^0}{\left|\mathbf{q}\right|-q^0}-i\pi \right]\right)
\end{aligned}
\end{equation}
with $m_D(T)$ the Debye mass of the gauge boson ($=\sqrt{2}g_3 T$ for the gluon).

With the above replacements, we can now perform the collision integrals numerically. We do so using the phase space parametrization discussed and detailed in Refs.~\cite{Moore:2001fga, Arnold:2002zm, Arnold:2003zc} and the \texttt{Vegas} Monte Carlo routine included in the \texttt{Cuba} package \cite{Hahn:2004fe}. The integrals converge reasonably quickly for most cases on a standard desktop computer to reasonable numerical precision.

The results of this numerical evaluation are given in Equations~\ref{eq:Hrate}--\ref{eq:bkg} below. These values can then be plugged into Eqs.~\ref{eq:Boltz} and \ref{eq:bkg_Boltz} for the perturbations. For the Higgs bosons, the rates (in the leading log approximation) are
\begin{equation}\label{eq:Hrate}
\begin{aligned}
&\Gamma_{\mu 1,h}^h\simeq(1.1\times 10^{-3}  g_3^2 y_t^2+6.0\times 10^{-4} y_t^4)T \\
& \Gamma_{T 1,h}^h\simeq  \Gamma_{\mu 2,h}^h\simeq(2.5\times 10^{-3}  g_3^2 y_t^2+1.4\times 10^{-3} y_t^4)T\\
& \Gamma_{T 2,h}^h\simeq(8.6\times 10^{-3}  g_3^2 y_t^2+4.8\times 10^{-3} y_t^4)T\\
& \Gamma_{v,h}^h\simeq(3.5\times 10^{-3}  g_3^2 y_t^2+1.8\times 10^{-3} y_t^4)T,
\end{aligned}
\end{equation}
while the corresponding contributions to the top quark distributions are
\begin{equation}\label{eq:Hratet}
\begin{aligned}
&-\Gamma_{\mu 1,t}^h\simeq(1.0\times 10^{-3}  g_3^2 y_t^2+5.8\times 10^{-4} y_t^4)T \\
& -\Gamma_{T 1,t}^h\simeq  \Gamma_{\mu 2,t}^h\simeq(2.5\times 10^{-3}  g_3^2 y_t^2+1.5\times 10^{-3} y_t^4)T\\
& -\Gamma_{T 2,t}^h\simeq(8.5\times 10^{-3}  g_3^2 y_t^2+4.8\times 10^{-3} y_t^4)T\\
& -\Gamma_{v,t}^h\simeq(2.8\times 10^{-3}  g_3^2 y_t^2+1.4\times 10^{-3} y_t^4)T.
\end{aligned}
\end{equation}
For the top quarks,
\begin{equation}\label{eq:trate}
\begin{aligned}
&\Gamma_{\mu 1,t}^t\simeq(5.0\times 10^{-4}  g_3^4+ 5.8\times 10^{-4}  g_3^2y_t^2+1.5\times 10^{-4} y_t^4)T \\
&\Gamma_{T 1,t}^t\simeq  \Gamma_{\mu 2,t}^t\simeq(1.2\times 10^{-3}  g_3^4+ 1.4\times 10^{-3}  g_3^2y_t^2+3.6\times 10^{-4} y_t^4)T\\
&\Gamma_{T 2,t}^t\simeq(1.1\times 10^{-2}  g_3^4+ 4.6\times 10^{-3}  g_3^2y_t^2+1.1\times 10^{-3} y_t^4)T\\
&\Gamma_{v,t}^t\simeq(2.0\times 10^{-2}  g_3^4+ 1.7\times 10^{-3}  g_3^2y_t^2+4.3\times 10^{-4} y_t^4)T,
\end{aligned}
\end{equation}
while their contributions to the Higgs distributions are
\begin{equation}\label{eq:trateH}
\begin{aligned}
&-\Gamma_{\mu 1,h}^t\simeq(9.3\times 10^{-5}  g_3^2y_t^2+5.3\times 10^{-5} y_t^4)T \\
&-\Gamma_{T 1,h}^t\simeq  \Gamma_{\mu 2,h}^t\simeq( 2.2\times 10^{-4}  g_3^2y_t^2+1.3\times 10^{-4} y_t^4)T\\
&-\Gamma_{T 2,h}^t\simeq(7.2\times 10^{-4}  g_3^2y_t^2+4.0\times 10^{-4} y_t^4)T\\
&-\Gamma_{v,h}^t\simeq(2.4\times 10^{-4}  g_3^2y_t^2+1.2\times 10^{-4} y_t^4)T.
\end{aligned}
\end{equation}
Finally, the background contributions are
\begin{equation}\label{eq:bkg}
\begin{aligned}
&\widetilde{ \Gamma}_{\mu 2,t}\simeq( 1.4\times 10^{-2} g_3^4+1.3\times 10^{-2}  g_3^2y_t^2+2.6\times 10^{-3} y_t^4)T\\
&\widetilde{\Gamma}_{T 2,t}\simeq( 1.4\times 10^{-1}  g_3^4+4.6\times 10^{-2} g_3^2y_t^2 + 8.7\times 10^{-3}y_t^4)T\\
&\widetilde{\Gamma}_{v,t}\simeq (2.4\times 10^{-1}  g_3^4 + 1.7 \times 10^{-2} g_3^2 y_t^2 +3.4\times 10^{-3} y_t^4)T\\
&\widetilde{\Gamma}_{\mu2,h}\simeq 0\\
&\widetilde{\Gamma}_{T2,h}\simeq( 1.0\times 10^{-3}  g_3^2y_t^2+9.8\times 10^{-5} y_t^4)T\\
&\widetilde{\Gamma}_{v,h}\simeq( 1.6\times 10^{-3}  g_3^2y_t^2+4.6\times 10^{-4} y_t^4)T.
\end{aligned}
\end{equation}
Note that the background rates for the top quarks tend to appear larger than their counterparts in Eq.~\ref{eq:trate} above. This is simply because in the background rates we have summed over all contributions, while the rates in Eqs.~\ref{eq:Hrate}--\ref{eq:trateH} are the average values per degree of freedom (e.g. divided by $N_t=6$ in the top quark case). The latter rates will be multiplied by the appropriate $N_i$ factors when they enter the bubble wall equation of motion.   Also, note that the contributions in Eqs.~\ref{eq:Hratet} and~\ref{eq:trateH} are negative because they arise from diagrams with the relevant species on the outgoing legs of the Feynman diagrams.

In previous work, the above integrals were performed analytically using several approximations and without incorporating the (momentum-dependent) self-energies. The different computational methods used here change the rates by $\mathcal{O}(1)$ factors relative to the results in Ref.~\cite{ Moore:1995si}. There were also some algebraic errors in the results of Ref.~\cite{ Moore:1995si}, as pointed out in Ref.~\cite{Arnold:2000dr}, that contribute to the discrepancy. Although our treatment is still formally at the same order as that of Ref.~\cite{ Moore:1995si}, the HTL-improved calculation in many cases is expected to more closely reproduce the leading order result (see e.g. Fig.~1 of Ref.~\cite{Arnold:2003zc}). Nevertheless, modulo the algebraic mistakes in Ref.~\cite{Moore:1995si}, our interaction rates are no more accurate than those of Moore and Prokopec in approximating the full leading order results; they should simply be thought of as arising from a different set of approximations. We comment further on the differences between the rates found in Ref.~\cite{Moore:1995si} and those reported above in Appendix~\ref{ap:rates}. The reader should bear in mind that the predicted wall velocity will tend to be higher if the rates computed in Ref.~\cite{Moore:1995si} are used instead of ours. This is because the former are larger and thus result in faster equilibration for the various perturbations. Note also that the collision integrals computed and listed above depend only on the Standard Model degrees of freedom, and as such are quite general. They can be used in various applications beyond those considered in this work. 

Before moving on, some comments regarding the higher order contributions neglected in the fluid approximation are in order.  The assumed form for the perturbations is that of a perfect fluid and can be thought of as a truncated expansion in powers of momentum.  That is, the fluid ansatz assumes that the effects of higher angular moments $p^{\ell}Y_{\ell m}(\widehat{\mathbf{p}})$ in the distribution functions are negligible \cite{Moore:1995si} (the $Y_{\ell m}$ are spherical harmonics).  For the top quarks this is a good approximation, since we find that the velocity perturbations typically satisfy $\delta v T / \delta \mu \lesssim 0.1$, while the contributions from higher moments should scale roughly  like $(\delta v T/\delta \mu)^{\ell}$ \cite{Moore:1995si}. On the other hand, the Higgs bosons have smaller interaction rates than the tops, and so the fluid approximation begins to break down for strong phase transitions. For this reason we will restrict ourselves to moderately strong phase transitions with $\phi_h(T_n)/T_n \lesssim 1.1$ in our consideration of the xSM in section~\ref{sec:resxSM}. Already in this regime some points will be found to possess no subsonic solutions. Further details on the limitations of the fluid approximation can be found in Appendix~B of Ref.~\cite{Moore:1995si}. 

\subsubsection{Singlet Contributions} \label{sec:singlet}

Excitations of the singlet field will also contribute to the friction on the bubble wall. The corresponding collision integral for singlet quasiparticles is dominated by scattering processes involving four external scalars. At high temperatures, the resulting effective interaction rates are typically suppressed relative to those for the processes involving external fermions. To see this, note that processes with $t$-channel diagrams involving two external scalars and two external fermions schematically contribute 
\begin{equation}
\Gamma_{\mu,1}\sim g^4 T \times \int_{m(T)}^{T} dq \frac{1}{q}
\end{equation}
to the first moment of the Boltzmann equation in the small-$q$ limit (here $q\equiv \left| \mathbf{p}-\mathbf{p}^{\prime}\right|$). The logarithmic divergence is cut off by the thermal self-energy of the exchanged quasiparticle. The upper limit $q\sim T$ corresponds to the breakdown of the small $q$ approximation. For processes involving four external scalars with a scalar exchanged in the $t$-channel, the integrals of the Bose-Einstein distribution functions are also IR sensitive. Cutting off the distribution functions with a parameter $\epsilon$ with mass dimension 1 such that $f_0(p/T)\rightarrow f_0(p/T+\epsilon/T)$, we find the corresponding contribution to be
\begin{equation} \label{eq:scalar}
\Gamma_{\mu,1}\sim \frac{a^4 T}{\epsilon ^2} \times \int_{m(T)}^{T} dq \frac{1}{q^3}
\end{equation}
in the small $q$ and $\epsilon/T$ regime ($a$ is a cubic coupling with mass dimension 1). The divergence is now nominally quadratic (again cut off by the self-energy of the exchanged scalar), and the integral of over the distribution functions is cut off by the thermal masses of the external scalars in the infrared. This suggests that a rough leading order estimate of the scalar quasiparticle scattering rates should include thermal masses in the Bose-Einstein distribution functions. Computing the dominant contributions involving the cubic and quartic couplings and performing the resulting integrals, we find interaction rates that are significantly smaller than those for the tops and Higgs across the range of couplings and temperatures we consider, despite the nominally more severe divergence structure. This is because for large temperatures, $\epsilon \sim g T$ and so the schematic rate in Eq.~\ref{eq:scalar} is suppressed by $1/T^3$. This is expected, since at high temperatures all dimensionful parameters of the zero temperature theory should be irrelevant \cite{Jeon:1994if}. Meanwhile the quartic interactions do not contribute a small $q$ divergence at leading order. We therefore expect the singlet qusiparticle collision term to be small. 

Assuming this is the case, the fluid approximation is likely to provide a rather poor estimate of the corresponding friction. Instead, we will make a `free particle' approximation \cite{Dine:1992wr, Liu:1992tn} for these excitations, dropping the collision term for the singlet. In this case, the Boltzmann equation can be solved exactly, without taking moments. The solution is given by Eq.~5.3 of Ref.~\cite{Moore:1995si}, to lowest order in $v_w$, and is reproduced below in Eq.~\ref{eq:exact}. We will include the corresponding contribution to the equations of motion when computing the friction. Note that the presence of non-negligible interactions would decrease the friction and increase the predicted wall velocity.

This treatment assumes that the dominant singlet excitations are well described by our perturbative kinetic theory, with a local collision term. This should be true for the hard excitations. For much softer excitations, a classical treatment with the short wavelength fluctuations integrated out is likely more appropriate \cite{Moore:1996bn, Moore:2000wx}. Unlike classical Yang-Mills fields (discussed in the next subsection), the classical scalar field is not overdamped \cite{Moore:2000mx}, and so infrared excitations are likely to equilibrate quickly. We thus neglect the effect of infrared singlet modes on the friction. This approximation is likely rather crude and should be revisited in the future. Including the IR contributions would increase the friction and potentially slow the wall down. The reader should bear this in mind as we proceed.

\subsection{Gauge Boson Contributions}\label{sec:gauge}

Finally, for our calculations incorporating the finite-temperature gauge boson cubic term in Eq.~\ref{eq:cubic}, we will include the friction from the $SU(2)_L$ gauge bosons. In contrast with the top quarks, Higgs, and singlet excitations, the friction in this case is dominated by infrared degrees of freedom, which can be treated approximately as over-damped classical fields \cite{Moore:2000mx} as opposed to the perturbative approach utilized for the other species. The distribution functions in the classical limit can be shown to satisfy \cite{Moore:2000mx}
\begin{equation} \label{eq:IR}
\frac{\pi m_{D,W}^2(T)}{8p} \frac{d f_W(p, T)}{d t}=-[p^2+m_W^2(\phi_h)] f_W(p,T) + \mathcal{N}
\end{equation} 
where $\mathcal{N}$ is a noise term. This equation serves as the analog of Eq.~\ref{eq:dtBoltz}. Further discussion of its derivation and applicability can be found in Ref.~\cite{Moore:2000mx}.

Note that hard gauge boson excitations also exert a drag force on the bubble wall, as computed in Ref.~\cite{Moore:1995si}. However, we have verified that these contributions are substantially suppressed relative to that from the IR gauge bosons, as found in Ref.~\cite{Moore:2000mx}. We do not include them.

\section{Solving for the Wall Velocity}\label{sec:solve}

With the collision terms evaluated, we can now solve the Boltzmann equations to determine the perturbations for a given field profile and wall velocity. The goal is then to find the value of $v_w$ and the profile (and hence the perturbations) such that the equations of motion are satisfied. We will describe how this can be done below. First, let us consider solutions to the Boltzmann equations given a particular profile and value of $v_w$.

\subsection{Exact Solution for the Singlet Excitations}\label{sec:singlet_sol}

As mentioned above, the singlet equation can be decoupled from the rest of the system and solved exactly in the free-particle limit. The result been discussed in detail previously \cite{Dine:1992wr, Liu:1992tn, Moore:1995si}, and so we simply quote it here. The integral appearing in the equations of motion~\ref{eq:wall_eom_1}, to lowest order in $v_w$, is
\begin{equation} \label{eq:exact}
\int \frac{d^3 p}{(2\pi)^3 2 E} \delta f_s(p,x)=v_w \int \frac{d^3 p}{(2\pi)^3 2 E} \frac{e^{E_p/T}}{\left(e^{E_p/T}\pm 1 \right)^2}\frac{ \mathcal{Q}(p_z)}{T}
\end{equation}
where the upper (lower) sign is for fermions (bosons). The function $\mathcal{Q}$ is defined as
\begin{equation}
\mathcal{Q}(p_z) =\left\{ \begin{array}{l r}
\sqrt{p_z^2+m_s(\phi_h,\phi_s,T)^2}-p_z, & p_z>-\sqrt{m_s^0(T)^2-m_s(\phi_h,\phi_s,T)^2} \\
-\sqrt{p_z^2+m_s(\phi_h,\phi_s,T)^2-m_s^0(T)^2} - p_z, \hspace{0.3cm} & p_z<-\sqrt{m_s^0(T)^2-m_s(\phi_h,\phi_s,T)^2}
\end{array} \right.
\end{equation}
with $m_s^0(T)$ the singlet mass (including thermal contributions) in the broken phase (i.e. at $z\rightarrow \infty$). This integral is multiplied by $\partial m_s^2(\phi_h,\phi_s,T)/\partial \phi_h$ in the Higgs EOM, and by $\partial m_s^2(\phi_h,\phi_s,T)/\partial \phi_s$ in the singlet equation.

\subsection{Exact Solution for the IR Gauge Contributions}

We can also solve for the classical gauge boson contribution in Eq.~\ref{eq:IR}. The result is \cite{Moore:2000mx}
\begin{equation} \label{eq:IRfrict}
\frac{d m_W^2(\phi_h)}{d\phi_h} \int \frac{d^3 p}{(2\pi)^3 2 E} \delta f_W(p,x)= v_w \frac{3 T}{32\pi} m_{D,W}^2(T)\frac{ \phi_h^{\prime}(x)}{\phi_h(x)^2} \hspace{0.1cm} \Theta (x-x_{*})
\end{equation}
where the quantity $x_*$ solves $m_W[\phi_h(x_*)]=1/L_h$, with $L_h$ the SM-like Higgs wall width. For smaller $x$, the WKB description used to derive Eq.~\ref{eq:IR} breaks down. For more discussion on this point, see Ref.~\cite{Moore:2000mx}. Note that this value cuts off the IR divergence of Eq.~\ref{eq:IRfrict}.

\subsection{Solving the top-Higgs System}

It remains to solve the equations for the top quark, Higgs, and background excitations. Here we follow the methods found in Refs.~\cite{Moore:1995si, John:2000zq} with some slight modifications.

Since we are interested in static solutions to the equations of motion in the wall frame,  all quantities depend only on $x$ and so the derivatives in Eqs.~\ref{eq:Boltz} can be re-written as $\partial_t \rightarrow v_w d/dx$, $\partial_z\rightarrow d/dx$. The Boltzmann equations in the static limit are therefore a set of linear ordinary inhomogeneous differential equations.

To solve them, the equations for the background temperature and velocity, Eqs.~\ref{eq:bkg_Boltz}, can be used to eliminate $T_{\rm bg}$ and $v_{\rm bg}$ from the top quark and Higgs equations. Defining a vector of perturbations
\begin{equation}
\bm{\delta}^{\rm T}\equiv \left(\delta \mu_t, \delta T_t, \delta v_t, \delta \mu_h, \delta T_h, \delta v_h \right),
\end{equation}
Eqs.~\ref{eq:Boltz} can then be written as
\begin{equation}
A_{lk} \frac{d}{d x} \delta_k + \Gamma_{lk}\delta_k=F_l
\end{equation}
with the definitions
\begin{equation}  
A \equiv \left(\begin{array}{c c} A_{tt} &0 \\ 0& A_{h h}
\end{array}\right), \hspace{0.3cm}
 \Gamma \equiv \left(\begin{array}{c c} \Delta_{tt} & \Delta_{t h} \\ \Delta_{h t} & \Delta_{h h}
\end{array}\right),
\hspace{0.3cm}
A_i\equiv \left(
\begin{array}{c c c}
v_w c_2^i&v_w c_3^i & \frac{1}{3}c_3^i \\
v_w c_3^i&v_w c_4^i & \frac{1}{3}c_4^i \\
\frac{1}{3}c_3^i & \frac{1}{3} c_4^i & \frac{1}{3} v_w c_4^i
\end{array}
\right),
\end{equation}
\vspace{0.1cm}
\begin{equation}
\Delta_{i j}\equiv\left(\begin{array}{c c c}
\vspace{0.1cm} \Gamma_{\mu 1, j}^i+\frac{c_3^i}{\overline{c}_4}\widetilde{\Gamma}_{\mu 2,j} \hspace{0.1cm} & \Gamma_{T 1, j}^i+\frac{c_3^i}{\overline{c}_4}\widetilde{\Gamma}_{T 2,j}  & 0 \\
\vspace{0.1cm}\Gamma_{\mu 2, j}^i+\frac{c_4^i}{\overline{c}_4}\widetilde{\Gamma}_{\mu 2,j} & \Gamma_{T 2, j}^i+\frac{c_4^i}{\overline{c}_4}\widetilde{\Gamma}_{T 2,j} & 0 \\
\vspace{0.1cm} 0&0& \hspace{-.1cm}T \Gamma_{v, j}^i+\frac{c_4^i T}{\overline{c}_4}\widetilde{\Gamma}_{v,j} 
\end{array}
\right)
\end{equation}
and the source vector
\begin{equation} 
\begin{aligned}
&\mathbf{F}(x)^{\rm T}\equiv \frac{v_w}{2 T} \left(c_1^t \frac{d m_t^2(\phi_h) }{dx}, \hspace{0.1cm} c_2^t \frac{d m_t^2(\phi_h)}{dx},  \hspace{0.1cm}  0,\hspace{0.1cm}  c_1^h(x) \frac{d m_h^2(\phi_h,\phi_s)}{dx} ,\hspace{0.1cm}   c_2^h \frac{d m_h^2(\phi_h,\phi_s)}{dx}, \hspace{0.1cm}  0\right).
\end{aligned}
\end{equation}
The field-dependent masses are given by Eq.~\ref{eq:masses}.

The system of equations can be solved by simple Green's function techniques. Following Ref.~\cite{Moore:1995si} we define the matrix $\chi$ such that
\begin{equation}
\left(A^{-1}\Gamma\right)_{ij}\chi_{jk}=\chi_{ik}\lambda_k
\end{equation}
where $\lambda_k$ are the eigenvalues of $A^{-1}\Gamma$. It is then straightforward to define the vector Green's function
\begin{equation}
G_i(x,y)=\operatorname{sgn}(\lambda_i)e^{-\lambda_i \left(x-y\right)}\Theta\left[\operatorname{sgn}(\lambda_i)(x-y)\right]
\end{equation}
in terms of which the perturbation $\delta_i$ is given by
\begin{equation}
\delta_i(x)=\chi_{ij}\int_{-\infty}^{\infty}\left[\chi^{-1}A^{-1} \mathbf{F}(y)\right]_jG_j(x,y)dy.
\end{equation}
These solutions can then be inserted into the equation for the variation in the background temperature,
\begin{equation}
\begin{aligned}
&\delta T_{\rm bg}(x)=\frac{1}{\overline{c}_4\left(\frac{1}{3}-v_w^2\right)} \int_{-\infty}^{x} \sum_i \left[T \widetilde{\Gamma}_{v,i}\delta v_i-v_w \left(\widetilde{\Gamma}_{\mu 2,i}\delta \mu_i+\widetilde{\Gamma}_{T2,i}\delta T_i \right)\right]\\
\end{aligned}
\end{equation}
defined with respect to $T_+$, its value far ahead of the bubble in the shock front.

\subsection{Approximate Solutions to the Equations of Motion} \label{sec:EOMsol}

With the perturbations determined, we can now try to identify solutions to the wall equations of motion.
In terms of the perturbations $\delta _j$, Eq.~\ref{eq:wall_eom_1} reads, for the gauge-invariant case,
\begin{equation} \label{eq:wall_eom}
\begin{aligned}
-(1-v_w^2) \phi_i^{\prime \prime}  + \frac{\partial V(\phi_i,T)}{\partial \phi_i} +&\sum_j \frac{\partial m_j^2(\phi_i)}{\partial \phi_i} \frac{T}{2}\left[c_{1}^j\delta \mu_j +c_{2}^j(\delta T_j +\delta T_{bg})\right] \\ &+ \frac{\partial m_s^2(\phi_i)}{\partial \phi_i} \int \frac{d^3p}{(2\pi)^3 2E} \delta f_s(x,p) = 0
\end{aligned}
\end{equation}
where the last term is given in Eq.~\ref{eq:exact}.  If the gauge boson contributions are included, the RHS of Eq.~\ref{eq:IRfrict} should be added to the LHS of the above expression. The boundary conditions are $\phi_{h,s}(x\rightarrow \mp \infty)=\phi_{h,s; \pm}(T_+)$ and $\phi_{h,s}^{\prime}(x\rightarrow \pm \infty)=0$. This system of equations will typically admit a solution for certain values of $v_w$ and profile $\phi_{h,s}(x)$.  Our strategy will be to vary the profile and scan over values of $v_w$ consistent with a deflagration bubble, looking for parameters such that the equations of motion (and Boltzmann equations) are simultaneously satisfied.  All parameter space points we consider have at most one deflagration solution.

Eq.~\ref{eq:wall_eom} represents a set of integro-differential equations for $\mathbf{\Phi}\equiv (\phi_h,\phi_s)^{\rm T}$ (in the following discussion it will be useful to use explicit vector notation).  To find its approximate solutions, we will follow the strategy of Refs.~\cite{Moore:1995si, John:2000zq} and use an ansatz for the field profiles which depend on only a few parameters.  Of course in using an ansatz it is unlikely that the full equations of motion will be satisfied exactly.  However, we can reasonably approximate a solution by scanning over the ansatz parameters and imposing physical constraints. For a given choice of parameters, the Boltzmann equations can be solved exactly and the results inserted into the EOM.  A set of parameter values such that all constraints are simultaneously satisfied corresponds to an approximate solution to the original equations. This strategy has been employed in previous calculations of the wall velocity \cite{Moore:1995ua,  Moore:1995si,  Konstandin:2014zta} and we expect the results obtained in this way to be a decent approximation to the full numerical solution.

Before analyzing Eq.~\ref{eq:wall_eom} further, some useful insight can be gained from solving the corresponding field equations with the assumption of constant friction of the form
\begin{equation} \label{eq:const_frict}
-(1-v_w^2) \frac{d^2 \mathbf{\Phi}}{dx^2}+ \nabla_{\phi} V(\mathbf{\Phi},T) + \mathcal{F}\frac{d \mathbf{\Phi}}{dx}= 0
\end{equation}
subject to the boundary conditions $\Phi_i(x\rightarrow \mp \infty)= \Phi_{i, \pm}$, $\Phi_i^{\prime} (x\rightarrow \pm \infty)=0$, where $\nabla_{\phi} \equiv (\partial/\partial\phi_h, \partial/\partial \phi_s)^{\rm T}$ and $\mathcal{F}$ is the same for both field directions. Clearly this is not a realistic case, but we will improve on it below.  These equations are much simpler than the full integro-differential equations of Eq.~\ref{eq:wall_eom}. As detailed in Ref.~\cite{Kozaczuk:2014kva}, the solution to the equations of motion can be found numerically via path deformations, and corresponds to a limit in which all the friction on the wall is parallel to the trajectory in the $(\phi_h,\phi_s)$ field space within the wall. This can be seen by noting that $d\mathbf{\Phi}/dx$ is a ``velocity'' vector in field space, so $\mathcal{F} d\mathbf{\Phi}/dx$ always acts parallel to the trajectory. In all cases we consider, the solution to these equations of motion is well-fit by a tanh ansatz with parameters
\begin{equation}\label{eq:ansatz}
\phi_i(x)=\frac{\phi_{i}^0}{2}\left(1+\tanh\frac{x-\delta_i}{L_i}\right)
\end{equation}
in the fluid frame. Here $L_i=L_h$, $L_s$ are the wall widths and $\delta_i$ are offsets allowing for a good fit to the numerical solution. We can take $\delta_h=0$ without loss of generality. Note that if we had allowed $\phi_s\neq 0$ in the electroweak-symmetric phase, we could have instead used $\phi_s(x)=\phi_s^0+\Delta \phi_{s}/2\left(1+\tanh\frac{x-\delta_s}{L_s}\right)$ for the ansatz, with $\phi_s^0$ the singlet VEV in the symmetric phase and $\Delta \phi_s$ the change in VEV across the wall. The remaining analysis would proceed in the same way.

How does the situation change when including a realistic friction term?  The friction is no longer proportional to $d\mathbf{\Phi}/dx$ and so will have some component perpendicular to the field space trajectory, acting as an effective ``normal force" along the path. However, there is a fortunate simplification we can make if the friction perpendicular to the field space trajectory found by solving Eq.~\ref{eq:const_frict} is negligible.  In this case, \emph{the field will not be significantly deformed from its field space path found using the constant friction equations of motion}, although the field profile in \emph{physical} space will change. In other words, if we write the solution to Eq.~\ref{eq:const_frict} as $\mathbf{\Phi}(s)$ where $s=s(x)$ is some parameter such that $\left|d \mathbf{\Phi}/ds\right|=1$, the effect of a change in the friction parallel to $\mathbf{\Phi}(s)$ will only be to alter $s(x)$. Meanwhile, a change in the friction normal to the profile would result in a change of $\mathbf{\Phi}(s)$ itself. Applying this reasoning to the tanh ansatz (which we find to be a good fit to the constant friction solution), the effect of altering only the friction parallel to the trajectory and neglecting that normal to the path will simply be an overall simultaneous re-scaling of all the widths and offsets: 
\begin{equation}
L_{h,s}\rightarrow a L_{h,s}, \hspace{0.5 cm} \delta_{s}\rightarrow a \delta_{s}.
\end{equation}
This is the only change in the tanh profile that will not deform the path in field space.  Then, starting from the constant friction solution, the problem can be reduced to finding the values of $v_w$ and $a$ such that the pressure and pressure gradient in the wall vanish:
\begin{equation} \begin{aligned} \label{eq:constraints}
&\int dx~ \left({\rm Eq.~\ref{eq:wall_eom}} \right) \cdot \frac{d \mathbf{\Phi}}{d x}=0,\\
&\int dx~ \left({\rm Eq.~\ref{eq:wall_eom}} \right) \cdot \frac{d^2 \mathbf{\Phi}}{d x^2}=0.
\end{aligned}
\end{equation} The above constraints will only be satisfied for values of $v_w$ and $a$ such that the wall is not accelerating or expanding/contracting, as required for the steady-state solution we are seeking.  This is a simple generalization of the strategy used in the SM case in Refs.~\cite{Moore:1995ua,  Moore:1995si, Konstandin:2014zta}, where $v_w$ and $L_w$ are varied. 

In what follows, we will assume that the friction force normal to the field space path determined from Eq.~\ref{eq:const_frict} is negligible, such that the discussion of the above paragraph applies. The validity of this assumption can be checked {\it a posteriori} (which we do), but there is an intuitive reason why it should often be reasonable.  At very high temperatures the potential is stabilized at the origin (in our approximation) by the effective thermal masses $\sim g T$ of all the scalars. Around $T\sim 0$, the potential is necessarily stabilized at a minimum away from the origin. The temperature of the phase transition is such that two minima exist simultaneously and are (nearly) degenerate. Provided that the tree-level contribution to the barrier is not too large, this can only occur if there is a significant cancellation between the zero-temperature and finite-temperature corrections in some direction of the $\mathbf{\Phi}$ field space. This approximate cancellation will be largest along the field space trajectory $\mathbf{\Phi}(s)$ found by solving Eq.~\ref{eq:const_frict} such that, schematically,
\begin{equation}\label{eq:cancel}
 \nabla_{\phi} V(\mathbf{\Phi},T=0)\cdot \frac{d \mathbf{\Phi}(s)}{d s} \sim - \sum \frac{dm(\mathbf{\Phi})^2}{d \mathbf{\Phi}(s)}\int \frac{d^3 p}{(2\pi)^3 2E}f_0(p,T)\cdot  \frac{d \mathbf{\Phi}(s)}{ds}.
 \end{equation}
The resulting ridge in the finite-temperature effective potential is precisely that along which the cubic term becomes relevant. We can insert the solution to Eq.~\ref{eq:const_frict} into Eq.~\ref{eq:wall_eom} and see how we expect the solution to change when going to the full EOM. With the approximate cancellation of Eq.~\ref{eq:cancel} in effect, the full equation of motion parallel to $\mathbf{\Phi}(s)$ is then schematically 
\begin{equation}
\left\{-(1-v_w^2)\frac{d^2 \mathbf{\Phi}}{dx^2}+ \sum \frac{dm(\mathbf{\Phi})^2}{d \mathbf{\Phi}}\int \frac{d^3 p}{(2\pi)^3 2E}\delta f(p,T) +\mathcal{O}(\delta f^2) \right\} \cdot \frac{d\mathbf{\Phi}(s)}{d s} \sim 0
 \end{equation}
 along $\mathbf{\Phi}(s)$.
The (approximate) cancellation has made the contribution from the friction the leading effect. The change in the friction term from Eq.~\ref{eq:const_frict} to~\ref{eq:wall_eom} will alter $s(x)$ but leave the field space trajectory $\mathbf{\Phi}(s)$ unchanged. In contrast, the cancellation in Eq.~\ref{eq:cancel} is not expected to hold in the perpendicular direction along the path ($\propto d^2\mathbf{\Phi}/ds^2$), unless the minimum of the potential away from the origin lies in the bottom of a shallow valley. In the absence of such an approximate continuous symmetry, the full EOM perpendicular to $\mathbf{\Phi}(s)$ is 
\begin{equation}
\left\{-(1-v_w^2) \frac{ d^2 \mathbf{\Phi}}{dx^2}+\nabla_{\phi} V(T=0)+ \sum \frac{dm(\mathbf{\Phi})^2}{d \mathbf{\Phi}}\int \frac{d^3 p}{(2\pi)^3 2E}f_0(p,T) +\mathcal{O}(\delta f)\right\} \cdot \frac{d^2\mathbf{\Phi}(s)}{d s^2} \sim 0
\end{equation}
and so the effect of the friction in this direction is perturbatively small, resulting in a negligible correction to the perpendicular component of Eq.~\ref{eq:const_frict} and hence to $\mathbf{\Phi}(s)$.

The above discussion suggests the following strategy for finding approximate solutions to the equations of motion for a given parameter space point:
\begin{enumerate}
\item Compute the phase transition properties, namely the order parameter and $T_n$. We do this using the \texttt{CosmoTransitions} package~\cite{Wainwright:2011kj}.
\item Solve for the constant friction profile from Eq.~\ref{eq:const_frict}. This can be done using path deformations \cite{Kozaczuk:2014kva} or otherwise. Fit the solution to the tanh ansatz Eq.~\ref{eq:ansatz}.
\item Solve the hydrodynamic relations to obtain $T_+$ for various values of $v_w$.
\item Vary the values of $v_w$ and $a$. For each pair, solve the Boltzmann equations as discussed above, using $T=T_+$ and $L_i=a L_i^0$, $\delta_s=a\delta_s^0$ with $L_i^0$ and $\delta_s^0$ obtained from the numerical solution of Eq.~\ref{eq:const_frict}.
\item Insert the solutions for the perturbations (and background temperature profile) into Eq.~\ref{eq:wall_eom}, then compute the constraints in Eq.~\ref{eq:constraints}. The values of $v_w$ and $a$ satisfying Eq.~\ref{eq:constraints} can be found by interpolating between the results of the scan.
\end{enumerate}
This method generalizes that of Refs.~\cite{Moore:1995ua,  Moore:1995si, Konstandin:2014zta} to accommodate the additional singlet field direction.

The results for $v_w$ and $a$ obtained in this way will still produce a residual `normal force' perpendicular to the trajectory in field space when inserted back into Eq.~\ref{eq:wall_eom} due to the neglect of the friction in the direction $\propto d^2 \mathbf{\Phi}/ds^2$. Defining $s(x)=\left|\mathbf{\Phi}(x)\right|$, the tangent and normal unit vectors to the field space path $\mathbf{\Phi}(s)$ are
\begin{equation}
\begin{aligned}
\hat{\mathbf{t}}(s)&=\frac{d \mathbf{\Phi}(s)}{ds} \hspace{0.1cm} \left|\frac{d \mathbf{\Phi}(s)}{ds}\right|^{-1}, \hspace{0.3cm} \hat{\mathbf{n}}(s)&=\frac{d}{ds}\hat{\mathbf{t}}(s).
\end{aligned}
\end{equation}
The `normal force' along $\mathbf{\Phi}(s)$ is given by
\begin{equation}\label{eq:N}
\mathbf{N}(x)=\frac{d^2\mathbf{\Phi}}{ds^2} \left(\frac{ds}{dx}\right)^2-\left[\nabla_{\phi}V(\mathbf{\Phi},T)+ \sum \frac{dm(\mathbf{\Phi})^2}{d \mathbf{\Phi}}\int \frac{d^3 p}{(2\pi)^3 2E}\delta f(p,T) \right]\cdot \hat{n}(s).
\end{equation}
A full solution to the equations of motion would guarantee that $\mathbf{N}(x)=0$ for all $x$. This will not be true for our approximate solutions. 

To check that this residual normal force is indeed negligible, we can deform the profile to eliminate it. The deformation can be performed along the lines suggested by Ref.~\cite{Wainwright:2011kj} for computing the critical bubble profile. It typically results in small changes to the original field space profile, which in turn have very little effect on the perturbations and constraints in Eq.~\ref{eq:constraints}, since the curvature perpendicular to the path is typically significant. This suggests that the wall velocity and profile found in the way outlined above should indeed provide a reasonable approximation to those obtained from the full solution of the equations of motion, at least in the cases we consider.  There may be exceptions elsewhere in the parameter space. 

Note that the procedure outlined in Steps 1-5 above is quite general, and can be adapted beyond singlet models to other scenarios with multiple field directions, provided Eq.~\ref{eq:cancel} is approximately satisfied. 

\section{Wall Velocities in the Real Singlet Extension}\label{sec:resxSM}

\subsection{Parameter Space and Phenomenology}\label{sec:pheno}
We now turn to the parameter space of the real singlet extension of the Standard Model as an application. Our goal is not a comprehensive analysis of this model. Instead, we focus on a sample of the parameter space consistent with current observations and a strongly first-order phase transition. 

The authors of Ref.~\cite{Profumo:2014opa} recently performed a detailed analysis of the electroweak phase transition in this setup and so we use their study as a guide. Recall that we have identified the excitation $h$ as the Standard Model-like Higgs with $m_h\simeq 125$ GeV. The couplings of the discovered Higgs are very close to those expected in the Standard Model~\cite{ATLAS_couplings, CMS_couplings}. Therefore we will assume that the doublet $H$ couples precisely as in the Standard Model and that there is no mixing between the singlet and Higgs at tree-level\footnote{Departing from this choice should not significantly affect our predicted range of wall velocities, since we drop the finite-temperature tadpole and thus the contribution from $a_1$.}. This corresponds to the choice
\begin{equation}\label{eq:nomix}
a_1+2a_2 v_s = 0
\end{equation}
and immediately fixes $\lambda$ in terms of the observed Higgs mass, $m_h^2=2\lambda v^2=(125$ GeV$)^2$. Note that Ref.~\cite{Profumo:2014opa} showed that deviations from this no-mixing limit are allowed by current LHC measurements and so this requirement can be relaxed. In our scans, we will vary both the cross-quartic coupling, $a_2$, and the zero-temperature singlet VEV, $v_s$. Then Eq.~\ref{eq:nomix} can be used to determine $a_1$, given our choice for $v_s$ and $a_2$. 

We will also assume $b_3=0$. The barrier required for a first-order electroweak phase transition will then arise primarily from the tree-level mixed cubic coupling $a_1$. Again, this is not required by the phenomenology, and this choice can be altered without significantly affecting any of our arguments. Note that much of the NMSSM parameter space compatible with a strongly first-order electroweak phase transition lies close to the corresponding limit $\left|\kappa A_{\kappa}\right| \ll \left| \lambda A_{\lambda}\right|$ \cite{Huang:2014ifa, Kozaczuk:2014kva}.

 Equating the minima of the tree-level potential with the VEVs $v=246$ GeV and $v_s$ yields the conditions
 \begin{equation}
 \begin{aligned} \mu^2&=\lambda v^2+\frac{1}{2}(a_1+a_2 v_s) v_s\\
 b_2&=-b_3 v_s-b_4 v_s^2-\frac{a_1 v^2}{4 v_s}-\frac{1}{2}a_2 v^2
 \end{aligned}
 \end{equation}
 which allows us to solve for $\mu^2$ and $b_2$. We also take the tree-level singlet mass squared (in the zero mixing limit),
 \begin{equation}\label{eq:b4}
 m_s^2=b_3 v_s+2b_4 v_s^2-\frac{a_1 v^2}{4 v_s}
 \end{equation} 
 as an input, and use the above expression to solve for $b_4$. 
 
 The free parameters are thus $a_2$, $v_s$, and $m_s^2$. There are some additional requirements on the theory that allow us to hone in on phenomenologically viable values for these quantities. First of all, stability of the $T=0$ potential requires $b_4>0$, which in turn limits the values of $m_s^2$ we can consider via Eq.~\ref{eq:b4}. Also, if $m_s^2$ is too small, the decay $h\rightarrow ss$ would cause large deviations in the width of $h$ which are not observed experimentally. On the other hand, if $m_s>2 m_h$, di-Higgs production at the LHC can place significant constraints on the model~\cite{No:2013wsa, Chen:2014ask}. For simplicity, we avoid this regime and choose $m_h/2<m_s<2m_h$.
 
 Further insight can be gained from considering the expected behavior of the electroweak phase transition strength and wall velocity as a function of these free parameters.  In particular, larger values of $a_2$ are favorable from the standpoint of small wall velocities. This is because increasing $a_2$ corresponds to lowering $T_c$, as can be seen by noting that $T_c$ corresponds approximately to the temperature such that $m^2_h(\phi_h, \phi_s, T)\sim 0$. Lowering $T_c$ results in lower values of $\phi_c$ satisfying $\phi_c/T_c\gtrsim 1$. Smaller field values in turn lower the total pressure difference between the vacua. This pressure difference is what drives the expansion of the bubble and which must be compensated for by the friction. Furthermore, larger values for $a_2$ correspond to larger $m_s^2(\phi_h,\phi_s)$, which in turn increases the friction from the singlet on the wall.

With the above reasoning in mind, we choose two sets of parameters likely to be promising for electroweak baryogenesis and across which we can compute the bubble wall velocity. These are 
\begin{equation}
\begin{aligned}
&\rm{Set \hspace{0.1cm}1:}\hspace{0.3 cm} m_s=170 \hspace{0.1cm} {\rm GeV},\hspace{0.1cm} a_2= 0.9 \\
&\rm{Set \hspace{0.1cm}2:}\hspace{0.3 cm} m_s=245 \hspace{0.1cm} {\rm GeV},\hspace{0.1cm} a_2= 1.7. 
 \end{aligned}
 \end{equation}
 For both sets of points we vary $v_s$, which corresponds to varying the strength of the electroweak phase transition. This is clear, since higher values of $v_s$ correspond to larger $|a_1|$ via Eq.~\ref{eq:nomix} and hence a larger contribution to the barrier separating the electroweak minimum from the origin at finite temperature. We vary $v_s$ up to values such that either the fluid approximation breaks down, or subsonic solutions no longer exist. This corresponds roughly to values of $v_s$ between 30-100 GeV for Sets 1 and 2. Note that with our choices of parameters, every point in Sets 1 and 2 will satisfy all current phenomenological constraints. Specifically, the electroweak vacuum is absolutely stable for all points considered while the absence of $s-h$ mixing ensures that both $h$ and the new singlet-like state are compatible with current observations and limits.
  
 \subsection{Results}
 
 \begin{figure}[!t]
\centering
\includegraphics[width=.65\textwidth]{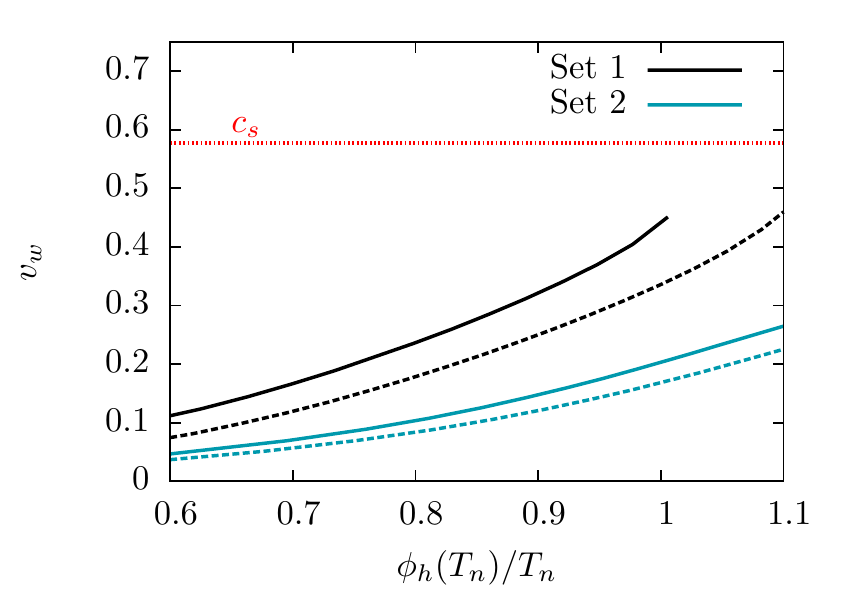}
\caption{\label{fig:results} Wall velocities for the xSM parameter space described in the text. The solid (dashed) curves depict the results neglecting (including) the $SU(2)_L$ gauge boson contributions to the finite temperature effective potential and friction. No subsonic solutions are found with $\phi_h(T_n)/T_n \gtrsim 1$ ($\gtrsim 1.1$) for the points in Set 1 neglecting (including) the gauge bosons.  The curves corresponding to Set 2 would extend beyond  $\phi_h(T_n)/T_n=1.1$, however the perturbative fluid approximation begins to break down significantly for stronger transitions, and so we restrict our results to the region shown. The red dotted line shows the speed of sound in the plasma, above which non-local electroweak baryogenesis is not possible. Note that we have searched exclusively for subsonic solutions to the equations of motion.}
\end{figure}

Finally we arrive at our results for the xSM. The wall velocities computed in a parameter scan for the two sets of points (Set 1 and 2) described above are shown in Fig.~\ref{fig:results}. The critical bubble profile and nucleation temperature are computed using \texttt{CosmoTransitions}~\cite{Wainwright:2011kj}.

The solid lines depict the outcome of the gauge-invariant method. Stronger transitions correspond to faster moving bubble walls. The perturbative fluid approximation becomes worse and breaks down for stronger phase transitions, and so we cut off our scans above $\phi_h(T_n)/T_n\sim 1.1$. Wall velocities for stronger phase transitions will only be larger than those shown. The dashed lines depict the resulting wall velocities when including the gauge boson cubic terms and friction. The values of $v_w$ are smaller in this case, although for Set 2 the gauge boson contribution makes a less significant difference. This suggests that our gauge-invariant treatment should provide a reasonable, though rough, estimate of the wall velocity when $v_w$ is not too large.

Fig.~\ref{fig:results} confirms our intuition from Sec.~\ref{sec:pheno}: larger thermal masses for the singlet and SM-like Higgs field result in slower bubble walls. Larger thermal masses trap the fields in the high-temperature minimum and delay the phase transition to lower temperatures. This yields smaller changes in the VEVs for a given phase transition strength, and hence a smaller pressure difference between the phases. The friction on the bubble wall also tends to be enhanced for larger thermal masses.

Interestingly, for strong first-order phase transitions, we find that subsonic solutions to the equations of motion may not exist. This is because as $v_w\rightarrow c_s$, the background temperature contribution begins to dominate in the Higgs and singlet field EOMs (it is proportional to $1/(c_s^2-v_w^2)$). As pointed out in Ref.~\cite{Konstandin:2014zta}, the background terms typically enter with a relative sign to those from the heavy species, thus \emph{reducing} the total friction for subsonic deflagrations.  This behavior is seen for Set 1  in Fig.~\ref{fig:results}: no subsonic solution exists for the gauge-invariant case with $\phi_h(T_n)/T_n \gtrsim 1$. Including the gauge-dependent terms, subsonic solutions can extend up to $\phi_h(T_n)/T_n \sim 1.1$, but not higher. We conclude that viable non-local electroweak baryogenesis in singlet-driven models is incompatible with very strong first-order phase transitions, at least in some cases. This can be at odds with sphaleron suppression inside the bubble, as seen for Set 1. 

Even if a subsonic solution exists, the bubbles tend to expand rather quickly from the standpoint of successful EWB. For example, previous studies of $CP$-violating sources in the MSSM~\cite{Huber:2001xf, Carena:2000id, Kozaczuk:2011vr} suggest that electroweak baryogenesis tends to be most efficient for $v_w\sim 0.01$, while Fig.~\ref{fig:results} indicates that  $v_w>0.2$ for most points featuring a strongly first-order phase transition. Viable bayogenesis in singlet-driven scenarios may thus require substantially more $CP$-violation than in models with slow walls (such as the MSSM with light stops) to overcome the suppression arising from large $v_w$.

 \begin{figure}[!t]
\centering
\includegraphics[width=.49\textwidth]{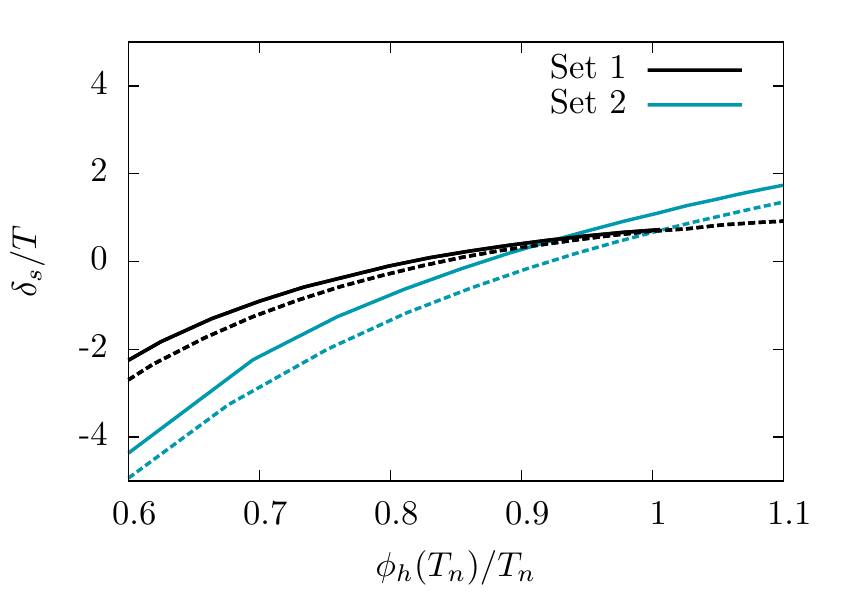}\\
\includegraphics[width=.49\textwidth]{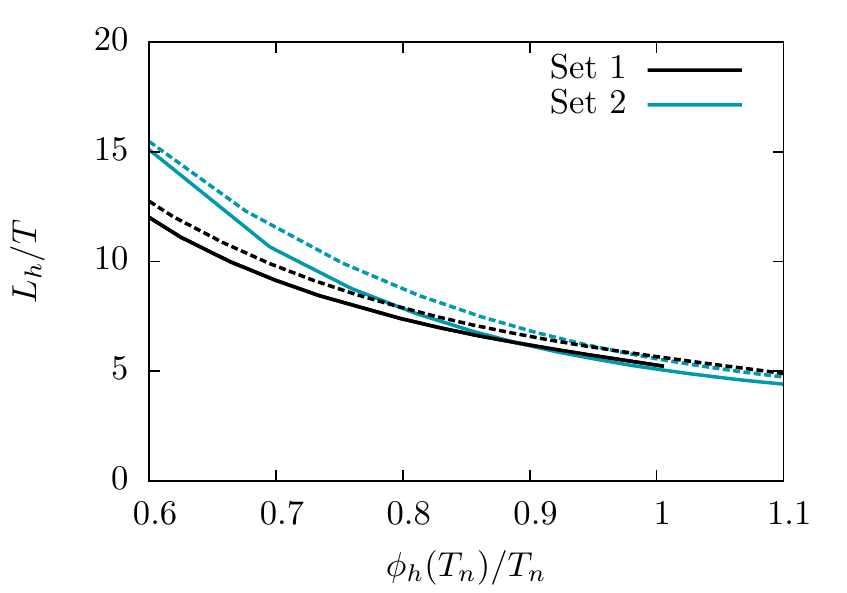} \includegraphics[width=.49\textwidth]{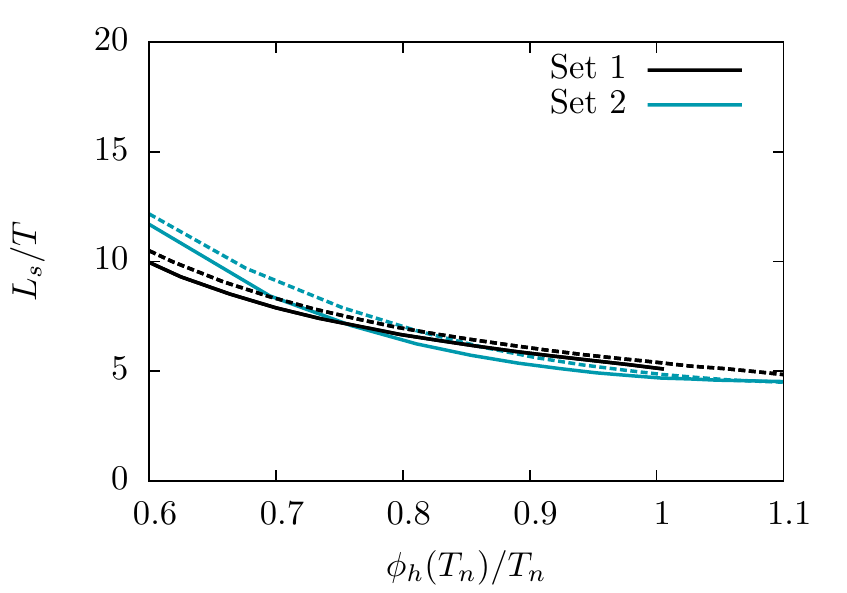}

\caption{\label{fig:profiles} Late-time bubble wall profiles relevant for electroweak baryogenesis obtained by solving the wall equations of motion. The solid (dashed) curves depict the results neglecting (including) the $SU(2)_L$ gauge boson contributions to the finite temperature effective potential and friction. The top panel shows the singlet field offset, while the bottom two show the SM-like Higgs and singlet wall widths. Bubbles with strong first-order phase transitions tend to feature $L_{h,s}\sim 5/T$ and the singlet lagging slightly behind  the Higgs field.}
\end{figure}

Our methods also allow us to determine the wall widths and offset for the subsonic configurations. These quantities are important inputs for microphysical calculations of the baryon asymmetry. The resulting bubble wall profiles for Sets 1 and 2 are shown in Fig.~\ref{fig:profiles}. The offset can change sign, with the singlet field lagging behind that of the SM-like Higgs for stronger phase transitions. For $\phi_h(T_n)/T_n\gtrsim 1$, the wall widths are typically $\sim \mathcal{O}(5/T)$. This is substantially smaller than typical values in Standard Model-like cases and consistent with the findings of Ref.~\cite{Kozaczuk:2014kva} in the NMSSM. Thin walls follow from the large pressure difference due to the changing singlet VEV during the transition. This is in fact promising for electroweak baryogenesis, since in many cases the $CP$-violating sources scale as $\sim 1/L_w$~\cite{Carena:2000id, Lee:2004we}.

One may ask to what extent we should expect similar results beyond the minimal real singlet extension of the Standard Model. After all, the xSM is known to be incomplete from the standpoint of electroweak baryogenesis, since it contains no new source of $CP$-violation. However, the model can be modified slightly to incorporate $CP$-violation by e.g. complexifying the singlet and adding $CP$-violating Higgs-singlet couplings, or by including additional higher dimension $CP$-violating operators, as in Ref.~\cite{Cline:2012hg}, in cases where the singlet VEV vanishes at $T=0$. Neither possibility should significantly alter the friction on the bubble wall. We expect similar conclusions in other $CP$-violating extensions of the xSM. Our findings followed primarily from the form of the friction, which is dominated by the top quarks and gauge bosons for the SM-like Higgs field and the singlet and Higgs excitations for the singlet field. As long as this is the case, the results beyond the minimal model should be qualitatively similar to those we have found here. In fact, this is not unreasonable: it would be difficult for new states to couple as strongly to the Higgs field as the top quark without violating existing phenomenological constraints, for example. Regardless, the methods and ingredients presented in sections~\ref{sec:Boltz}-\ref{sec:solve} can be used to determine the wall velocity beyond the minimal xSM, although this may require computing additional interaction rates involving the excitations of the new species in the plasma.

\section{Summary and Conclusions} \label{sec:summ}

In this study, we have seen how to compute the electroweak bubble wall velocity at singlet-driven first-order phase transitions. This extends previous work which applied to the Standard Model-- and MSSM--like cases. For concreteness, we framed our discussion in the real singlet extension of the Standard Model, or xSM, although our methods can be used in other models involving singlets at the electroweak phase transition.

Some of the key findings of this study are:

\begin{itemize}
\item As anticipated, bubbles tend to expand rather quickly at first-order phase transitions driven by tree-level cubic terms in which the singlet vacuum expectation value changes appreciably. We have found $v_w\gtrsim 0.2$ for all points with $\phi_h(T_n)/T_n \geq 1$ in the xSM. These wall velocities may be compatible with electroweak baryogenesis in some cases, provided a sufficiently strong source of $CP$-violation. One should bear in mind, however, that the free-particle approximation made for the singlet excitations may overestimate the corresponding friction, and thus lead to an underestimate of $v_w$.

\item The most promising parameter space for slower bubble walls, and hence for electroweak baryogenesis, features larger thermal masses for the singlet and SM-like Higgs field. This translates into larger values of $a_2$ and $b_4$ in the xSM.

\item Strong phase transitions may exhibit no subsonic solution and hence not allow for viable transport-driven electroweak baryogenesis. For example, considering a particular set of parameters in the xSM, we found no points with $v_w<c_s$ for $\phi_h(T_n)/T_n\gtrsim 1.1$.

\item A gauge-invariant estimation of the bubble wall velocity is possible and should approximate the full solution rather well for the slowest walls. The (gauge-dependent) $SU(2)_L$ gauge boson contributions become important for faster moving bubble walls.

\item Wall widths are typically of order $\sim 5/T$ for strong first-order phase transitions as required for electroweak baryogenesis. These values are considerably smaller than their Standard Model analogs which are often used in the literature.

\end{itemize}

Along with the above points, our treatment of the friction on the bubble wall can be useful in various applications related to the electroweak plasma. For example, the interaction rates computed for the top quarks and Higgs bosons can be used to extract diffusion constants for these species, valid at leading log order and including the effects of hard thermal loops, which are important in transport calculations for electroweak baryogenesis.  A rough estimate along the lines of Ref.~\cite{Joyce:1994zt, Moore:1995si} suggests $D_h\sim 13/T$, $D_t\sim 2/T$.

Our results are promising from the standpoint of observable gravitational radiation. The peak amplitude for the stochastic gravity wave background produced at a phase transition is enhanced for faster moving bubbles and larger pressure differences. The ingredients presented in this study can be used to more precisely compute the resulting gravity wave spectrum in concrete models involving singlets. Recent work~\cite{Hindmarsh:2013xza, Hindmarsh:2015qta} suggests that the peak amplitude of the signal from a strong electroweak-scale phase transition may in fact be significantly larger than previously realized. It would be interesting to analyze singlet driven phase transitions given these new hydrodynamic insights along with our predictions for the wall velocity in concrete models. An observable gravity wave signal could provide exciting (indirect) evidence for a first-order phase transition, and possibly electroweak baryogenesis, in the early Universe. 

There are several ways to improve over the methods presented this study. One might hope to move beyond the simple fluid approximation to be able to study stronger phase transitions. Also beneficial would be a full leading-order determination of the quasiparticle interaction rates entering the Boltzmann equations for the various perturbations. Other improvements include accounting for the effects of the spherical bubble geometry on the hydrodynamics, formulating a gauge-independent treatment incorporating the gauge and Goldstone bosons (which is a difficult problem), and considering full numerical solutions to the bubble wall equations of motion rather than utilizing an ansatz. These improvements, required for more precise determinations of the wall velocity, would become much more important if the LHC or a future collider were to unearth direct evidence for a singlet-extended Higgs sector. In the interim, we expect our methods to provide a decent approximation of the bubble wall velocity in singlet-driven scenarios, which remain a particularly compelling setting for electroweak baryogenesis in light of current experimental constraints.

\begin{acknowledgments}
\noindent  I would like to thank David E. Morrissey and Stefano Profumo for insightful conversations surrounding this project and for feedback on this manuscript, as well as Nikita Blinov for many useful discussions.  I am also grateful to the organizers and participants of the 2014 KITP `Particlegenesis' program, where I first began to think more carefully about this subject. This research was supported by the National Science and Engineering Research Council of Canada (NSERC).
\end{acknowledgments}

\appendix
\section{Comparing Effective Interaction Rates}\label{ap:rates}

The effective interaction rates we have computed and listed in Equations~\ref{eq:Hrate}--\ref{eq:bkg} differ by $\mathcal{O}(1)$ factors from those appearing in the classic references~\cite{Joyce:1994zn, Moore:1995si}. In this appendix, we show that this discrepancy can be explained by the slightly different set of approximations made in evaluating the integrals analytically in previous works. Technically, both the methods used in Refs.~\cite{Joyce:1994zn, Moore:1995si} and in this work are valid `leading log approximations' to the full first-order result, in that only processes contributing logarithmically to the collision integrals are considered. They differ primarily in their treatments of the non-logarithmic pieces of the various momentum integrals. The discrepancies can therefore be understood to demonstrate the uncertainties associated with the leading log approximation and the importance of performing a full leading-order calculation for more precise results in future studies. 

To understand the different approaches to evaluating the collision integrals, let us consider the process $t\bar{t}\rightarrow g g$. First of all, Ref.~\cite{Moore:1995si} does not include the symmetry factor for the corresponding matrix elements, resulting in a factor of two discrepancy before evaluating any integrals \cite{Arnold:2000dr}. Including the symmetry factor, the leading-log matrix element is $\approx 64/9 g_3^4 u/t$, where we have averaged over the top quark degrees of freedom. The resulting contribution to $\Gamma_{\mu 1, t}^t$, evaluating the integral numerically and including the top quark momentum-dependent self-energy, is
\begin{equation}
\Delta \Gamma_{\mu1,t}^t\approx1.1\times 10^{-3} T,
\end{equation}
whereas Ref.~\cite{Moore:1995si} reports
\begin{equation} \label{eq:MP}
\Delta \Gamma_{\mu1,t}^t\simeq \frac{16 \alpha_s^2}{9\pi^3}\times \frac{9 \zeta_2^2}{16}\log\frac{9T^2}{m_q^2} \hspace{0.1cm} T \approx3.8\times 10^{-3}T,
\end{equation}
including the correct symmetry factor. Starting from the same matrix element, the different methods for evaluating the integrals results in almost a factor of 4 difference in the result.  

Simply evaluating the integral in Ref.~\cite{Moore:1995si} numerically, but including the thermal mass in the propagator instead of the HTL momentum-dependent self-energy, we find
\begin{equation}
\Delta \Gamma_{\mu1,t}^t\approx 1.5\times 10^{-3} T,
\end{equation}
suggesting that the simple propagator replacement over-estimates the integral by about $40\%$, but does not account for the whole discrepancy. In fact, most of the difference comes from the various approximations made to arrive at the analytic result in Eq.~\ref{eq:MP}. In particular, all non-logarithmic contributions are dropped, while numerically evaluating the integrals includes all of the various contributions. 

For example, the final result for $\Delta \Gamma_{\mu1,t}^t$ in Ref.~\cite{Moore:1995si} contains an integral over the plasma frame angle $\theta$ between $\mathbf{p}$ and $\mathbf{k}$, 
\begin{equation}
\int d\cos\theta \frac{1}{2}\log \left(\frac{2 \left|\mathbf{p}\right|\left|\mathbf{k}\right|\left(1-\cos\theta\right)}{m_t^2}\right)=-1+\log\frac{4\left|\mathbf{p}\right|\left|\mathbf{k}\right|}{m_t^2}.
\end{equation}
Moore and Prokopec drop the constant piece, keeping only the logarithm. However, the contribution of the constant piece is numerically comparable, and of opposite sign, to the logarithmic term. Performing the remaining integrals over $\left|\mathbf{p}\right|$, $\left|\mathbf{k}\right|$, the contribution without the constant term is $\sim 3.1\times 10^{-3}T$, while including it yields $\sim 1.8\times10^{-3}T$, which is significantly closer to the results we have obtained. Similar approximations are made in performing the other integrals, and for the other rates. 

It is worth reiterating that neither approach includes all processes contributing at leading order in the gauge couplings. Our interaction rates should simply be viewed as a slightly different approximation to the full leading order result. Future studies of the wall velocity in different models may find it beneficial to compare the results obtained from our interaction rates and those of Ref.~\cite{Moore:1995si} to assess the uncertainty expected from the leading log approximation.


\begin{thebibliography}{300}

\bibitem{Turner:1990rc} 
  M.~S.~Turner and F.~Wilczek,
  Phys.\ Rev.\ Lett.\  {\bf 65}, 3080 (1990).
  
  \bibitem{Hogan:1986qda} 
  C.~J.~Hogan,
  Mon.\ Not.\ Roy.\ Astron.\ Soc.\  {\bf 218}, 629 (1986).
  
  \bibitem{Witten:1984rs} 
  E.~Witten,
  Phys.\ Rev.\ D {\bf 30}, 272 (1984).
  
\bibitem{Kamionkowski:1993fg} 
  M.~Kamionkowski, A.~Kosowsky and M.~S.~Turner,
  Phys.\ Rev.\ D {\bf 49}, 2837 (1994)
  [astro-ph/9310044].
    
    
  \bibitem{Turner:1987bw} 
  M.~S.~Turner and L.~M.~Widrow,
  Phys.\ Rev.\ D {\bf 37}, 2743 (1988).
  
  \bibitem{Baym:1995fk} 
  G.~Baym, D.~Bodeker and L.~D.~McLerran,
  Phys.\ Rev.\ D {\bf 53}, 662 (1996)
  [hep-ph/9507429].
  
  \bibitem{Cohen:2008nb} 
  T.~Cohen, D.~E.~Morrissey and A.~Pierce,
  Phys.\ Rev.\ D {\bf 78}, 111701 (2008)
  [arXiv:0808.3994 [hep-ph]].
  
   \bibitem{Wainwright:2009mq} 
  C.~Wainwright and S.~Profumo,
  Phys.\ Rev.\ D {\bf 80}, 103517 (2009)
  [arXiv:0909.1317 [hep-ph]].
  
  \bibitem{Boyanovsky:2006bf} 
  D.~Boyanovsky, H.~J.~de Vega and D.~J.~Schwarz,
  Ann.\ Rev.\ Nucl.\ Part.\ Sci.\  {\bf 56}, 441 (2006)
  [hep-ph/0602002].
  
  \bibitem{Coleman:1977py} 
  S.~R.~Coleman,
  Phys.\ Rev.\ D {\bf 15}, 2929 (1977)
  [Phys.\ Rev.\ D {\bf 16}, 1248 (1977)].
  
  \bibitem{Callan:1977pt} 
  C.~G.~Callan, Jr. and S.~R.~Coleman,
  Phys.\ Rev.\ D {\bf 16}, 1762 (1977).
  
  \bibitem{Linde1} 
  A.~D.~Linde,
  Phys.\ Lett.\ B {\bf 100}, 37 (1981).
  
  \bibitem{Linde2} 
  A.~D.~Linde,
  Nucl.\ Phys.\ B {\bf 216}, 421 (1983)
  [Nucl.\ Phys.\ B {\bf 223}, 544 (1983)].

  \bibitem{Quiros_review} 
  M.~Quiros,
  hep-ph/9901312.
  
   \bibitem{Patel:2011th} 
  H.~H.~Patel and M.~J.~Ramsey-Musolf,
  JHEP {\bf 1107}, 029 (2011)
  [arXiv:1101.4665 [hep-ph]].
  
  \bibitem{Fuyuto:2014yia} 
  K.~Fuyuto and E.~Senaha,
  Phys.\ Rev.\ D {\bf 90}, no. 1, 015015 (2014)
  [arXiv:1406.0433 [hep-ph]].
  
  \bibitem{Cohen:1990it} 
  A.~G.~Cohen, D.~B.~Kaplan and A.~E.~Nelson,
  Nucl.\ Phys.\ B {\bf 349}, 727 (1991).
  
  \bibitem{Cohen:1990py} 
  A.~G.~Cohen, D.~B.~Kaplan and A.~E.~Nelson,
  Phys.\ Lett.\ B {\bf 245}, 561 (1990).
  
  \bibitem{Nelson:1991ab} 
  A.~E.~Nelson, D.~B.~Kaplan and A.~G.~Cohen,
  Nucl.\ Phys.\ B {\bf 373}, 453 (1992).
  
  \bibitem{Joyce:1994zn} 
  M.~Joyce, T.~Prokopec and N.~Turok,
  Phys.\ Rev.\ D {\bf 53}, 2930 (1996)
  [hep-ph/9410281].
  
  \bibitem{Joyce:1994zt} 
  M.~Joyce, T.~Prokopec and N.~Turok,
  Phys.\ Rev.\ D {\bf 53}, 2958 (1996)
  [hep-ph/9410282].
  
  \bibitem{Morrissey:2012db} 
  D.~E.~Morrissey and M.~J.~Ramsey-Musolf,
  New J.\ Phys.\  {\bf 14}, 125003 (2012)
  [arXiv:1206.2942 [hep-ph]].
  
  \bibitem{Engel:2013lsa} 
  J.~Engel, M.~J.~Ramsey-Musolf and U.~van Kolck,
  Prog.\ Part.\ Nucl.\ Phys.\  {\bf 71}, 21 (2013)
  [arXiv:1303.2371 [nucl-th]].
  
  \bibitem{Joyce:1994fu} 
  M.~Joyce, T.~Prokopec and N.~Turok,
  Phys.\ Rev.\ Lett.\  {\bf 75}, 1695 (1995)
  [Phys.\ Rev.\ Lett.\  {\bf 75}, 3375 (1995)]
  [hep-ph/9408339].
  
  \bibitem{No:2011fi} 
  J.~M.~No,
  Phys.\ Rev.\ D {\bf 84}, 124025 (2011)
  [arXiv:1103.2159 [hep-ph]].
  
   \bibitem{Heckler:1994uu} 
  A.~F.~Heckler,
  Phys.\ Rev.\ D {\bf 51}, 405 (1995)
  [astro-ph/9407064].
 
  \bibitem{Huber:2001xf} 
  S.~J.~Huber, P.~John and M.~G.~Schmidt,
  Eur.\ Phys.\ J.\ C {\bf 20}, 695 (2001)
  [hep-ph/0101249].
  
  \bibitem{Carena:2000id} 
  M.~Carena, J.~M.~Moreno, M.~Quiros, M.~Seco and C.~E.~M.~Wagner,
  Nucl.\ Phys.\ B {\bf 599}, 158 (2001)
  [hep-ph/0011055].
  
  \bibitem{Kozaczuk:2011vr} 
  J.~Kozaczuk and S.~Profumo,
  JCAP {\bf 1111}, 031 (2011)
  [arXiv:1108.0393 [hep-ph]].
  
  \bibitem{Konstandin:2005cd} 
  T.~Konstandin, T.~Prokopec, M.~G.~Schmidt and M.~Seco,
  Nucl.\ Phys.\ B {\bf 738}, 1 (2006)
  [hep-ph/0505103].

  \bibitem{Huber:2006wf} 
  S.~J.~Huber, T.~Konstandin, T.~Prokopec and M.~G.~Schmidt,
  Nucl.\ Phys.\ B {\bf 757}, 172 (2006)
  [hep-ph/0606298].
 
  \bibitem{Huber:2006ma} 
  S.~J.~Huber, T.~Konstandin, T.~Prokopec and M.~G.~Schmidt,
  Nucl.\ Phys.\ A {\bf 785}, 206 (2007)
  [hep-ph/0608017].

  \bibitem{Kainulainen:2002th} 
  K.~Kainulainen, T.~Prokopec, M.~G.~Schmidt and S.~Weinstock,
  Phys.\ Rev.\ D {\bf 66}, 043502 (2002)
  [hep-ph/0202177].
  
  \bibitem{Prokopec:2003pj} 
  T.~Prokopec, M.~G.~Schmidt and S.~Weinstock,
  Annals Phys.\  {\bf 314}, 208 (2004)
  [hep-ph/0312110].
  
  \bibitem{Konstandin:2004gy} 
  T.~Konstandin, T.~Prokopec and M.~G.~Schmidt,
  Nucl.\ Phys.\ B {\bf 716}, 373 (2005)
  [hep-ph/0410135].
  
  \bibitem{Kajantie:1995kf} 
  K.~Kajantie, M.~Laine, K.~Rummukainen and M.~E.~Shaposhnikov,
  Nucl.\ Phys.\ B {\bf 466}, 189 (1996)
  [hep-lat/9510020].
  
  \bibitem{Kajantie:1996mn} 
  K.~Kajantie, M.~Laine, K.~Rummukainen and M.~E.~Shaposhnikov,
  Phys.\ Rev.\ Lett.\  {\bf 77}, 2887 (1996)
  [hep-ph/9605288].
  
  \bibitem{Pietroni:1992in} 
  M.~Pietroni,
  Nucl.\ Phys.\ B {\bf 402}, 27 (1993)
  [hep-ph/9207227].
 
  
  \bibitem{Davies:1996qn} 
  A.~T.~Davies, C.~D.~Froggatt and R.~G.~Moorhouse,
  Phys.\ Lett.\ B {\bf 372}, 88 (1996)
  [hep-ph/9603388].
  
  \bibitem{Huber:2000mg} 
  S.~J.~Huber and M.~G.~Schmidt,
  Nucl.\ Phys.\ B {\bf 606}, 183 (2001)
  [hep-ph/0003122].
  
  \bibitem{Menon:2004wv} 
  A.~Menon, D.~E.~Morrissey and C.~E.~M.~Wagner,
  Phys.\ Rev.\ D {\bf 70}, 035005 (2004)
  [hep-ph/0404184].
  
  \bibitem{Ham:2004cf} 
  S.~W.~Ham, Y.~S.~Jeong and S.~K.~Oh,
  J.\ Phys.\ G {\bf 31}, 857 (2005)
  [hep-ph/0411352].
  
  \bibitem{Profumo:2007wc} 
  S.~Profumo, M.~J.~Ramsey-Musolf and G.~Shaughnessy,
  JHEP {\bf 0708}, 010 (2007)
  [arXiv:0705.2425 [hep-ph]].
  
  \bibitem{Bodeker:2009qy} 
  D.~Bodeker and G.~D.~Moore,
  JCAP {\bf 0905}, 009 (2009)
  [arXiv:0903.4099 [hep-ph]].
  
  \bibitem{Espinosa:2011ax} 
  J.~R.~Espinosa, T.~Konstandin and F.~Riva,
  Nucl.\ Phys.\ B {\bf 854}, 592 (2012)
  [arXiv:1107.5441 [hep-ph]].
  
  \bibitem{Kozaczuk:2013fga} 
  J.~Kozaczuk, S.~Profumo and C.~L.~Wainwright,
  Phys.\ Rev.\ D {\bf 87}, no. 7, 075011 (2013)
  [arXiv:1302.4781 [hep-ph]].
  
  \bibitem{Huang:2014ifa} 
  W.~Huang, Z.~Kang, J.~Shu, P.~Wu and J.~M.~Yang,
  Phys.\ Rev.\ D {\bf 91}, no. 2, 025006 (2015)
  [arXiv:1405.1152 [hep-ph]].
  
  \bibitem{Kozaczuk:2014kva} 
  J.~Kozaczuk, S.~Profumo, L.~S.~Haskins and C.~L.~Wainwright,
  JHEP {\bf 1501}, 144 (2015)
  [arXiv:1407.4134 [hep-ph]].
  
  \bibitem{Profumo:2014opa} 
  S.~Profumo, M.~J.~Ramsey-Musolf, C.~L.~Wainwright and P.~Winslow,
  Phys.\ Rev.\ D {\bf 91}, no. 3, 035018 (2015)
  [arXiv:1407.5342 [hep-ph]].

  \bibitem{Curtin:2014jma} 
  D.~Curtin, P.~Meade and C.~T.~Yu,
  JHEP {\bf 1411}, 127 (2014)
  [arXiv:1409.0005 [hep-ph]].
  
  \bibitem{Jiang:2015cwa} 
  M.~Jiang, L.~Bian, W.~Huang and J.~Shu,
  arXiv:1502.07574 [hep-ph].
  
  \bibitem{Aad:2012tfa} 
  G.~Aad {\it et al.}  [ATLAS Collaboration],
  Phys.\ Lett.\ B {\bf 716}, 1 (2012)
  [arXiv:1207.7214 [hep-ex]].
  
  \bibitem{Chatrchyan:2012ufa} 
  S.~Chatrchyan {\it et al.}  [CMS Collaboration],
  Phys.\ Lett.\ B {\bf 716}, 30 (2012)
  [arXiv:1207.7235 [hep-ex]].
  
   \bibitem{Konstandin:2014zta} 
  T.~Konstandin, G.~Nardini and I.~Rues,
  arXiv:1407.3132 [hep-ph].
  
  \bibitem{Kosowsky:1991ua} 
  A.~Kosowsky, M.~S.~Turner and R.~Watkins,
  Phys.\ Rev.\ D {\bf 45}, 4514 (1992).
  

  \bibitem{Kosowsky:1992rz} 
  A.~Kosowsky, M.~S.~Turner and R.~Watkins,
  Phys.\ Rev.\ Lett.\  {\bf 69}, 2026 (1992).
  
  \bibitem{Kosowsky:1992vn} 
  A.~Kosowsky and M.~S.~Turner,
  Phys.\ Rev.\ D {\bf 47}, 4372 (1993)
  [astro-ph/9211004].
  
  \bibitem{Kosowsky:2001xp} 
  A.~Kosowsky, A.~Mack and T.~Kahniashvili,
  Phys.\ Rev.\ D {\bf 66}, 024030 (2002)
  [astro-ph/0111483].
  
  \bibitem{Caprini:2006jb} 
  C.~Caprini and R.~Durrer,
  Phys.\ Rev.\ D {\bf 74}, 063521 (2006)
  [astro-ph/0603476].
  
  \bibitem{Gogoberidze:2007an} 
  G.~Gogoberidze, T.~Kahniashvili and A.~Kosowsky,
  Phys.\ Rev.\ D {\bf 76}, 083002 (2007)
  [arXiv:0705.1733 [astro-ph]].
  
 
  \bibitem{Huber:2007vva} 
  S.~J.~Huber and T.~Konstandin,
  JCAP {\bf 0805}, 017 (2008)
  [arXiv:0709.2091 [hep-ph]].
  
  \bibitem{Caprini:2007xq} 
  C.~Caprini, R.~Durrer and G.~Servant,
  Phys.\ Rev.\ D {\bf 77}, 124015 (2008)
  [arXiv:0711.2593 [astro-ph]].
  
   
  \bibitem{Huber:2008hg} 
  S.~J.~Huber and T.~Konstandin,
  JCAP {\bf 0809}, 022 (2008)
  [arXiv:0806.1828 [hep-ph]].
  
  \bibitem{Caprini:2009fx} 
  C.~Caprini, R.~Durrer, T.~Konstandin and G.~Servant,
  Phys.\ Rev.\ D {\bf 79}, 083519 (2009)
  [arXiv:0901.1661 [astro-ph.CO]].
  
  \bibitem{Caprini:2009yp} 
  C.~Caprini, R.~Durrer and G.~Servant,
  JCAP {\bf 0912}, 024 (2009)
  [arXiv:0909.0622 [astro-ph.CO]].
  
  \bibitem{Leitao:2012tx} 
  L.~Leitao, A.~Megevand and A.~D.~Sanchez,
  JCAP {\bf 1210}, 024 (2012)
  [arXiv:1205.3070 [astro-ph.CO]].

  \bibitem{Hindmarsh:2013xza} 
  M.~Hindmarsh, S.~J.~Huber, K.~Rummukainen and D.~J.~Weir,
  Phys.\ Rev.\ Lett.\  {\bf 112}, 041301 (2014)
  [arXiv:1304.2433 [hep-ph]].
  
  \bibitem{Hindmarsh:2015qta} 
  M.~Hindmarsh, S.~J.~Huber, K.~Rummukainen and D.~J.~Weir,
  arXiv:1504.03291 [astro-ph.CO].
  
  \bibitem{Binetruy:2012ze} 
  P.~Binetruy, A.~Bohe, C.~Caprini and J.~F.~Dufaux,
  JCAP {\bf 1206}, 027 (2012)
  [arXiv:1201.0983 [gr-qc]].
  
  \bibitem{Corbin:2005ny} 
  V.~Corbin and N.~J.~Cornish,
  Class.\ Quant.\ Grav.\  {\bf 23}, 2435 (2006)
  [gr-qc/0512039].
  
  \bibitem{Enqvist:1991xw} 
  K.~Enqvist, J.~Ignatius, K.~Kajantie and K.~Rummukainen,
  Phys.\ Rev.\ D {\bf 45}, 3415 (1992).
  
  \bibitem{Dine:1992wr} 
  M.~Dine, R.~G.~Leigh, P.~Y.~Huet, A.~D.~Linde and D.~A.~Linde,
  Phys.\ Rev.\ D {\bf 46}, 550 (1992)
  [hep-ph/9203203].
  
  \bibitem{Liu:1992tn} 
  B.~H.~Liu, L.~D.~McLerran and N.~Turok,
  Phys.\ Rev.\ D {\bf 46}, 2668 (1992).
  
 \bibitem{Ignatius:1993qn} 
  J.~Ignatius, K.~Kajantie, H.~Kurki-Suonio and M.~Laine,
  Phys.\ Rev.\ D {\bf 49}, 3854 (1994)
  [astro-ph/9309059].
  
  \bibitem{Carrington:1993ng} 
  M.~E.~Carrington and J.~I.~Kapusta,
  Phys.\ Rev.\ D {\bf 47}, 5304 (1993).

  \bibitem{Moore:1995ua} 
  G.~D.~Moore and T.~Prokopec,
  Phys.\ Rev.\ Lett.\  {\bf 75}, 777 (1995)
  [hep-ph/9503296].
 
  \bibitem{Moore:1995si} 
  G.~D.~Moore and T.~Prokopec,
  Phys.\ Rev.\ D {\bf 52}, 7182 (1995)
  [hep-ph/9506475].
 
  
  \bibitem{John:2000zq} 
  P.~John and M.~G.~Schmidt,
  Nucl.\ Phys.\ B {\bf 598}, 291 (2001)
  [Erratum-ibid.\ B {\bf 648}, 449 (2003)]
  [hep-ph/0002050].
  
  \bibitem{Moore:2000wx} 
  G.~D.~Moore,
  JHEP {\bf 0003}, 006 (2000)
  [hep-ph/0001274].
  
\bibitem{Megevand:2009gh} 
  A.~Megevand and A.~D.~Sanchez,
  Nucl.\ Phys.\ B {\bf 825}, 151 (2010)
  [arXiv:0908.3663 [hep-ph]].

\bibitem{Huber:2011aa} 
  S.~J.~Huber and M.~Sopena,
  Phys.\ Rev.\ D {\bf 85}, 103507 (2012)
  [arXiv:1112.1888 [hep-ph]].
 
\bibitem{Huber:2013kj} 
  S.~J.~Huber and M.~Sopena,
  arXiv:1302.1044 [hep-ph].
  
  \bibitem{Espinosa:2010hh} 
  J.~R.~Espinosa, T.~Konstandin, J.~M.~No and G.~Servant,
  JCAP {\bf 1006}, 028 (2010)
  [arXiv:1004.4187 [hep-ph]].
  
  
  \bibitem{Leitao:2010yw} 
  L.~Leitao and A.~Megevand,
  Nucl.\ Phys.\ B {\bf 844}, 450 (2011)
  [arXiv:1010.2134 [astro-ph.CO]].

  \bibitem{Megevand:2013hwa} 
  A.~MŽgevand,
  JCAP {\bf 1307}, 045 (2013)
  [arXiv:1303.4233 [astro-ph.CO]].
  
  \bibitem{Megevand:2013yua} 
  A.~Megevand and F.~A.~Membiela,
  Phys.\ Rev.\ D {\bf 89}, no. 10, 103507 (2014)
  [arXiv:1311.2453 [astro-ph.CO]].
  
  \bibitem{Leitao:2014pda} 
  L.~Leitao and A.~Megevand,
  Nucl.\ Phys.\ B {\bf 891}, 159 (2015)
  [arXiv:1410.3875 [hep-ph]].
  
  \bibitem{Megevand:2014yua} 
  A.~Megevand and F.~A.~Membiela,
  Phys.\ Rev.\ D {\bf 89}, no. 10, 103503 (2014)
  [arXiv:1402.5791 [astro-ph.CO]].
  
  \bibitem{Megevand:2014dua} 
  A.~Megevand, F.~A.~Membiela and A.~D.~Sanchez,
  JCAP {\bf 1503}, no. 03, 051 (2015)
  [arXiv:1412.8064 [hep-ph]].
 
  \bibitem{Carena:1996wj} 
  M.~Carena, M.~Quiros and C.~E.~M.~Wagner,
  Phys.\ Lett.\ B {\bf 380}, 81 (1996)
  [hep-ph/9603420].
  
  \bibitem{Delepine:1996vn} 
  D.~Delepine, J.~M.~Gerard, R.~Gonzalez Felipe and J.~Weyers,
  Phys.\ Lett.\ B {\bf 386}, 183 (1996)
  [hep-ph/9604440].
  
  \bibitem{Delgado:2012eu} 
  A.~Delgado, G.~F.~Giudice, G.~Isidori, M.~Pierini and A.~Strumia,
  Eur.\ Phys.\ J.\ C {\bf 73}, no. 3, 2370 (2013)
  [arXiv:1212.6847 [hep-ph]].
  
  \bibitem{Krizka:2012ah} 
  K.~Krizka, A.~Kumar and D.~E.~Morrissey,
  Phys.\ Rev.\ D {\bf 87}, no. 9, 095016 (2013)
  [arXiv:1212.4856 [hep-ph]].
  
  \bibitem{Cohen:2012zza} 
  T.~Cohen, D.~E.~Morrissey and A.~Pierce,
  Phys.\ Rev.\ D {\bf 86}, 013009 (2012)
  [arXiv:1203.2924 [hep-ph]].
  
  \bibitem{Menon:2009mz} 
  A.~Menon and D.~E.~Morrissey,
  Phys.\ Rev.\ D {\bf 79}, 115020 (2009)
  [arXiv:0903.3038 [hep-ph]].
  
  \bibitem{Curtin:2012aa} 
  D.~Curtin, P.~Jaiswal and P.~Meade,
  JHEP {\bf 1208}, 005 (2012)
  [arXiv:1203.2932 [hep-ph]].
  
  \bibitem{Carena:2012np} 
  M.~Carena, G.~Nardini, M.~Quiros and C.~E.~M.~Wagner,
  JHEP {\bf 1302}, 001 (2013)
  [arXiv:1207.6330 [hep-ph]].
  
  \bibitem{Chung:2012vg} 
  D.~J.~H.~Chung, A.~J.~Long and L.~T.~Wang,
  Phys.\ Rev.\ D {\bf 87}, no. 2, 023509 (2013)
  [arXiv:1209.1819 [hep-ph]].
  
  \bibitem{Katz:2014bha} 
  A.~Katz and M.~Perelstein,
  JHEP {\bf 1407}, 108 (2014)
  [arXiv:1401.1827 [hep-ph]].
  
  \bibitem{Turok:1990in} 
  N.~Turok and J.~Zadrozny,
  Phys.\ Rev.\ Lett.\  {\bf 65}, 2331 (1990).
  
  \bibitem{McLerran:1990zh} 
  L.~D.~McLerran, M.~E.~Shaposhnikov, N.~Turok and M.~B.~Voloshin,
  Phys.\ Lett.\ B {\bf 256}, 451 (1991).
  
  \bibitem{Dine:1990fj} 
  M.~Dine, P.~Huet, R.~L.~Singleton, Jr and L.~Susskind,
  Phys.\ Lett.\ B {\bf 257}, 351 (1991).
  
  \bibitem{Lue:1996pr} 
  A.~Lue, K.~Rajagopal and M.~Trodden,
  Phys.\ Rev.\ D {\bf 56}, 1250 (1997)
  [hep-ph/9612282].

 \bibitem{Tranberg:2003gi} 
  A.~Tranberg and J.~Smit,
  JHEP {\bf 0311}, 016 (2003)
  [hep-ph/0310342].
  
  \bibitem{Tranberg:2009de} 
  A.~Tranberg, A.~Hernandez, T.~Konstandin and M.~G.~Schmidt,
  Phys.\ Lett.\ B {\bf 690}, 207 (2010)
  [arXiv:0909.4199 [hep-ph]].
  
  \bibitem{Konstandin:2011ds} 
  T.~Konstandin and G.~Servant,
  JCAP {\bf 1107}, 024 (2011)
  [arXiv:1104.4793 [hep-ph]].
  
  \bibitem{Caprini:2011uz} 
  C.~Caprini and J.~M.~No,
  JCAP {\bf 1201}, 031 (2012)
  [arXiv:1111.1726 [hep-ph]].

  \bibitem{O'Connell:2006wi} 
  D.~O'Connell, M.~J.~Ramsey-Musolf and M.~B.~Wise,
  Phys.\ Rev.\ D {\bf 75}, 037701 (2007)
  [hep-ph/0611014].
  
  \bibitem{Barger:2007im} 
  V.~Barger, P.~Langacker, M.~McCaskey, M.~J.~Ramsey-Musolf and G.~Shaughnessy, 
  Phys.\ Rev.\ D {\bf 77}, 035005 (2008)
  [arXiv:0706.4311 [hep-ph]].
  
  \bibitem{He:2009yd} 
  X.~G.~He, T.~Li, X.~Q.~Li, J.~Tandean and H.~C.~Tsai,
  Phys.\ Lett.\ B {\bf 688}, 332 (2010)
  [arXiv:0912.4722 [hep-ph]].
  
  \bibitem{Gonderinger:2009jp} 
  M.~Gonderinger, Y.~Li, H.~Patel and M.~J.~Ramsey-Musolf,
  JHEP {\bf 1001}, 053 (2010)
  [arXiv:0910.3167 [hep-ph]].
  
  \bibitem{Cline:2012hg} 
  J.~M.~Cline and K.~Kainulainen,
  JCAP {\bf 1301}, 012 (2013)
  [arXiv:1210.4196 [hep-ph]].
  
  \bibitem{Cline:2013gha} 
  J.~M.~Cline, K.~Kainulainen, P.~Scott and C.~Weniger,
  Phys.\ Rev.\ D {\bf 88}, 055025 (2013)
  [arXiv:1306.4710 [hep-ph]].

  \bibitem{Nielsen:1975fs} 
  N.~K.~Nielsen,
  Nucl.\ Phys.\ B {\bf 101}, 173 (1975).
  
  \bibitem{Fukuda:1975di} 
  R.~Fukuda and T.~Kugo,
  Phys.\ Rev.\ D {\bf 13}, 3469 (1976).

 \bibitem{Fuyuto:2014yia} 
  K.~Fuyuto and E.~Senaha,
  Phys.\ Rev.\ D {\bf 90}, no. 1, 015015 (2014)
  [arXiv:1406.0433 [hep-ph]].
  
  \bibitem{Konstandin:2010dm} 
  T.~Konstandin and J.~M.~No,
  JCAP {\bf 1102}, 008 (2011)
  [arXiv:1011.3735 [hep-ph]].
  
  \bibitem{Jeon:1995zm} 
  S.~Jeon and L.~G.~Yaffe,
  Phys.\ Rev.\ D {\bf 53}, 5799 (1996)
  [hep-ph/9512263].
  
  \bibitem{Arnold:2002zm} 
  P.~B.~Arnold, G.~D.~Moore and L.~G.~Yaffe,
  JHEP {\bf 0301}, 030 (2003)
  [hep-ph/0209353].
  
  \bibitem{Comelli:1996vm} 
  D.~Comelli and J.~R.~Espinosa,
  Phys.\ Rev.\ D {\bf 55}, 6253 (1997)
  [hep-ph/9606438].
  
  \bibitem{Arnold:2000dr} 
  P.~B.~Arnold, G.~D.~Moore and L.~G.~Yaffe,
  JHEP {\bf 0011}, 001 (2000)
  [hep-ph/0010177].

    \bibitem{Arnold:2003zc} 
  P.~B.~Arnold, G.~D.~Moore and L.~G.~Yaffe,
  JHEP {\bf 0305}, 051 (2003)
  [hep-ph/0302165].
  
  \bibitem{Jeon:1994if} 
  S.~Jeon,
  Phys.\ Rev.\ D {\bf 52}, 3591 (1995)
  [hep-ph/9409250].
  
 \bibitem{Klimov:1981ka} 
  V.~V.~Klimov,
  Sov.\ J.\ Nucl.\ Phys.\  {\bf 33}, 934 (1981)
  [Yad.\ Fiz.\  {\bf 33}, 1734 (1981)].
  
  \bibitem{Weldon:1982bn} 
  H.~A.~Weldon,
  Phys.\ Rev.\ D {\bf 26}, 2789 (1982).
  

  \bibitem{Moore:2001fga} 
  G.~D.~Moore,
  JHEP {\bf 0105}, 039 (2001)
  [hep-ph/0104121].
  
  \bibitem{Hahn:2004fe} 
  T.~Hahn,
  Comput.\ Phys.\ Commun.\  {\bf 168}, 78 (2005)
  [hep-ph/0404043].
  
  \bibitem{Moore:1996bn} 
  G.~D.~Moore and N.~Turok,
  Phys.\ Rev.\ D {\bf 55}, 6538 (1997)
  [hep-ph/9608350].
  
  \bibitem{Moore:2000mx} 
  G.~D.~Moore,
  Phys.\ Rev.\ D {\bf 62}, 085011 (2000)
  [hep-ph/0001216].
  
  \bibitem{Wainwright:2011kj} 
  C.~L.~Wainwright,
  Comput.\ Phys.\ Commun.\  {\bf 183}, 2006 (2012)
  [arXiv:1109.4189 [hep-ph]].
  
  \bibitem{ATLAS_couplings} 
  The ATLAS collaboration,
  ATLAS-CONF-2015-007, ATLAS-COM-CONF-2015-011.
 
  \bibitem{CMS_couplings} 
  V.~Khachatryan {\it et al.}  [CMS Collaboration],
  Eur.\ Phys.\ J.\ C {\bf 75}, no. 5, 212 (2015)
  [arXiv:1412.8662 [hep-ex]].
  
  \bibitem{No:2013wsa} 
  J.~M.~No and M.~Ramsey-Musolf,
  Phys.\ Rev.\ D {\bf 89}, no. 9, 095031 (2014)
  [arXiv:1310.6035 [hep-ph]].
  
  \bibitem{Chen:2014ask} 
  C.~Y.~Chen, S.~Dawson and I.~M.~Lewis,
  Phys.\ Rev.\ D {\bf 91}, no. 3, 035015 (2015)
  [arXiv:1410.5488 [hep-ph]].
  
  
  \bibitem{Lee:2004we} 
  C.~Lee, V.~Cirigliano and M.~J.~Ramsey-Musolf,
  Phys.\ Rev.\ D {\bf 71}, 075010 (2005)
  [hep-ph/0412354].
  
 \end{thebibliography}
\end{document}